\documentclass[a4paper,reprint,twocolumn,aps,pra, superscriptaddress,floatfix]{revtex4-2}

\usepackage{amssymb,amsmath,amsthm,bbm,bbold,ragged2e,multirow,tcolorbox,enumitem}
\usepackage{amsfonts}
\usepackage{graphicx}
\usepackage{mathtools}
\usepackage{natbib}
\usepackage[version=4]{mhchem}
\usepackage[normalem]{ulem}
\usepackage[dvipsnames]{xcolor}
\usepackage{subfigure}
\usepackage{soul}
\usepackage[breaklinks,colorlinks,allcolors=blue]{hyperref}
\usepackage{upgreek}

\usepackage{pgfplots}
\pgfplotsset{compat=1.18}
\usepackage{fontawesome5}
\usetikzlibrary{shadows, calc, arrows.meta, fit, backgrounds}

\newcommand{\rme}{\mathrm{e}}

\newcommand{\g}{\gamma}
\newcommand{\G}{\Gamma}
\newcommand{\Gmf}{\G_\mathrm{mf}^{(2)}}
\newcommand{\bra}[1]{\mbox{$\langle #1 |$}}
\newcommand{\ket}[1]{\mbox{$| #1 \rangle$}}
\newcommand{\F}{\mathcal{F}}
\newcommand{\Fc}{\mathcal{F}^\mathrm{c}}
\newcommand{\Fmf}{\mathcal{F}^\mathrm{mf}}

\newcommand{\Ecum}{E_{\mathrm{cum}}}

\newcommand{\Sph}{S_\mathrm{ph}}

\newcommand{\kc}{\kappa^\mathrm{c}}

\newcommand{\iDMFT}{\text{i-DMFT}} 
\DeclareMathOperator{\trace}{Tr}

\newcommand{\RDMset}{\mathcal{E}_N^1}
\newcommand{\tRDMset}{\mathcal{E}_N^2}
\renewcommand{\Re}{\mathop\mathsf{Re}}

\begin{document}

\author{Paul G. Graf}
\affiliation{Arnold Sommerfeld Centre for Theoretical Physics, Ludwig-Maximilians-Universität München, Munich, Germany}

\author{Florian Matz}
\affiliation{Arnold Sommerfeld Centre for Theoretical Physics, Ludwig-Maximilians-Universität München, Munich, Germany}

\author{Markus Penz}
\affiliation{Arnold Sommerfeld Centre for Theoretical Physics, Ludwig-Maximilians-Universität München, Munich, Germany}
\affiliation{Department of Computer Science, Oslo Metropolitan University, Oslo, Norway}

\author{Christian Schilling}
\email{christian.schilling@unige.ch}
\affiliation{Department of Physical Chemistry, University of Geneva, Geneva, Switzerland}
\affiliation{Arnold Sommerfeld Centre for Theoretical Physics, Ludwig-Maximilians-Universität München, Munich, Germany}

\title{Putting a Temperature on Electron Correlation
}

\begin{abstract}
It is common wisdom that a better understanding of how to quantify electron correlation should lead to more efficient treatments of the correlation problem. Here, we not only substantiate this view but turn it into a concrete, systematic framework for ground states. Since prominent correlation measures such as the nonfreeness depend only on the one-particle reduced density matrix, modeling the correlation energy as a function of such a measure naturally places the framework within reduced density matrix functional theory, while keeping the cost at mean-field level. Strikingly, the variable conjugate to the measure emerges as a ``correlation temperature'', thereby connecting correlation measures, functional theory, and effective temperatures. The framework rationalizes methods such as \iDMFT{}, and we illustrate its key concepts analytically for the Hubbard dimer and numerically for small molecular systems. This brings a fundamental question within reach: how much of the correlation problem can be solved at mean-field cost?
\end{abstract}

\maketitle

\section{Introduction}
\label{sec:intro}

Describing electron correlation is one of the central challenges of quantum chemistry, condensed matter physics, and materials science, particularly for strongly correlated systems~\cite{Ganoe-Shee-2024}. Its difficulty stems from the exponential complexity of the many-electron wave function~\cite{Lowdin1958,davidson-book}. In quantum chemistry, the most common starting point for tackling this correlation problem is the Hartree--Fock (HF) method. By restricting the wave function to a single Slater determinant, HF reduces the problem to self-consistent one-particle equations at so-called mean-field cost. Moreover, HF is rarely the end point: perturbative, coupled-cluster, configuration-interaction, and many multireference approaches typically build on the HF reference state and its orbitals. HF thus constitutes the natural first step for most quantum-chemical treatments of electron correlation.

A remarkable insight due to Lieb~\cite{LiebHF,Lieb1977} is that, for positive interactions, the HF method can be recast as the minimization of a functional of the one-particle reduced density matrix (1RDM) over all ensemble $N$-representable 1RDMs. HF thereby becomes the simplest approximation within reduced density matrix functional theory (RDMFT)~\cite{Gilbert75,Valone80,M84,Goedecker1998,PG15}, whose exact functional differs from the HF functional only by the correlation functional $\Fc$. Despite well-developed foundations~\cite{Altunbulak2008,Baldsiefen2015,Schilling18,GR19,liebert23,Fredheim2025}, however, approximations to $\Fc$ are still less mature than in density functional theory (DFT), and their systematic improvement remains a central challenge.

Quantum information theory offers a complementary perspective, starting from a more basic question: what is electron correlation in the first place? Since electrons do not constitute proper subsystems in the operator-algebraic sense, the standard paradigm based on local operations and classical communication does not apply, and no unique measure of electron correlation exists~\cite{AlivertiPiuri2024}. Instead, it can only be quantified ad hoc, for instance by the prominent nonfreeness, the distance of a state from the set of mean-field states~\cite{Gottlieb2005,GottliebMauser2014}. Remarkably, for pure states the nonfreeness is a simple function of the 1RDM. Moreover, these mean-field states are precisely those whose two-particle reduced density matrix underlies the HF functional. Both perspectives thus single out the 1RDM and rest on the same notion of uncorrelated states.

\begin{figure}[t]
    \centering
    \begin{tikzpicture}[
        font=\sffamily,
        medal/.style={
            circle,
            minimum size=2.2cm,
            line width=5pt,
            fill=white,
            align=center,
            drop shadow={shadow xshift=1.5mm, shadow yshift=-1.2mm, opacity=.35}
        },
        lbl/.style={
            align=center,
            font=\footnotesize\bfseries,
            text=black!70,
            text width=2.8cm
        },
        biarrow/.style={
            {Latex[length=2.6mm,width=2mm]}-{Latex[length=2.6mm,width=2mm]},
            line width=1.1pt,
            opacity=.85
        }
    ]
    \colorlet{cmeas}{teal!50!blue}
    \colorlet{cfunc}{violet!50!purple}
    \colorlet{ctemp}{orange!50!red}
    %
    \coordinate (top)   at (0, 1.6);
    \coordinate (left)  at (-1.8,-1.3);
    \coordinate (right) at ( 1.8,-1.3);
    
    \node[medal, draw=cmeas, text=cmeas] (Ic) at (top)
        {\huge\faRuler\\[.3em]\Large$\zeta(\g)$};
    
    \node[medal, draw=cfunc, text=cfunc] (Wc) at (left)
        {\huge\faCogs\\[.3em]\Large$\tilde\Fc(\zeta)$};
    
    \node[medal, draw=ctemp, text=ctemp] (Tc) at (right)
        {\huge\faThermometerHalf\\[.3em]\Large$\kc$};
    
    \node[lbl,color=cmeas] (labIc) at ($(top)+(0,1.8)$)
        {correlation measure};
    
    \node[lbl,color=cfunc] (labWc) at ($(left)+(0,-1.8)$)
        {correlation functional};
    
    \node[lbl,color=ctemp!80] (labTc) at ($(right)+(0,-1.8)$)
        {correlation temperature};
    
    \draw[biarrow, draw=ctemp!50!cmeas] (Tc) to[bend left=10] (Ic);
    \draw[biarrow, draw=cmeas!50!cfunc] (Ic) to[bend left=10] (Wc);
    \draw[biarrow, draw=cfunc!50!ctemp] (Wc) to[bend left=10] (Tc);
    
    \begin{scope}[on background layer]
    \node[
        draw=black,
        line width=1.0pt,
        rounded corners=9pt,
        fill=white,
        drop shadow={shadow xshift=1.5mm, shadow yshift=-1.2mm, opacity=.28},
        fit=(Tc)(Ic)(Wc)(labTc)(labIc)(labWc),
        inner ysep=10pt,
        inner xsep=18pt
    ] {};
    \end{scope}
    
    \end{tikzpicture}
    \caption{Three concepts for modeling electron correlation connected by the framework developed in this work: a correlation measure $\zeta(\g)$ of the 1RDM, the coarse-grained correlation functional $\tilde\Fc(\zeta)$, and the correlation temperature $\kc$ (see text for details).}
    \label{fig:concepts}
\end{figure}
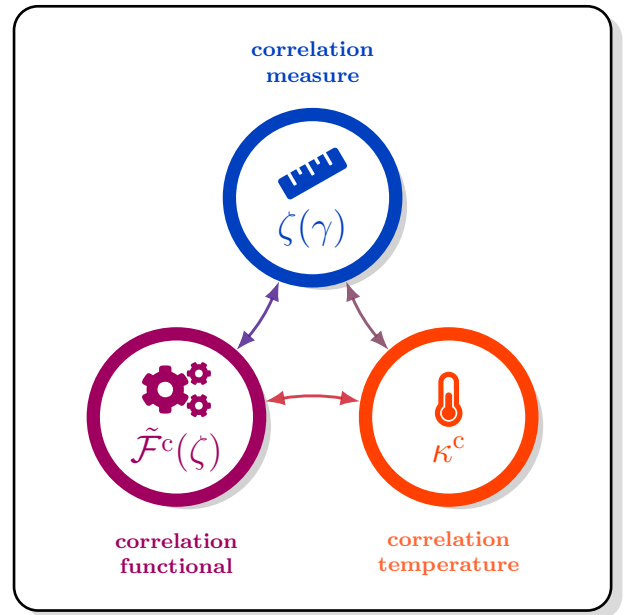

This convergence lends weight to the common wisdom that a better understanding of how to quantify electron correlation should lead to more efficient treatments of the correlation problem. Yet it leaves open how a correlation measure could actually be put to work in a computational method. In this perspective, we not only substantiate this view but turn it into a concrete, systematic framework for ground states. The guiding idea is simple: if a correlation measure captures the essential correlation of the $N$-electron wave function, as compressed into its 1RDM, then the correlation energy should be expressible through the measure alone, at least within a specific class of systems, such as a molecule along its dissociation coordinate. Following this idea through naturally places the framework within RDMFT, while keeping the computational cost at the mean-field level. Most importantly, it reveals that three seemingly distinct concepts are intimately connected: correlation measures from quantum information theory, RDMFT, and effective temperatures, whose thermal occupation numbers mimic those of correlated states (Fig.~\ref{fig:concepts}). This connection, a central result of this work, gives such temperatures a clear conceptual origin: unlike the occupation smearing used to improve numerical convergence~\cite{Weinert1992,Rabuck1999}, they follow directly from quantifying electron correlation.

Seen from a different angle, our work addresses a question that, for strongly correlated systems, has received comparatively little attention: rather than gaining accuracy by investing ever more computational resources, how far beyond HF can one go at fixed mean-field cost?
Several methods already explore this territory by introducing fractional occupations through an effective temperature, among them \iDMFT{}~\cite{WangKnowles21-PRA,WangBaerends22-PRL}, thermally-assisted-occupation DFT (TAO-DFT)~\cite{Chai2012,Chai2014,Yeh2022}, and entropy-corrected pair-density functional theory~\cite{Feng2025}. Our framework provides these approaches with a common foundation, which clarifies where they work and how they could be improved systematically.

The perspective is organized as follows. Section~\ref{sec:corr-measures} reviews correlation measures based on the 1RDM and formulates the conditions they should satisfy. Section~\ref{sec:corr-temp} develops the framework for constructing functional approximations and introduces the correlation temperature. Section~\ref{sec:Hubbard} illustrates the key concepts analytically for the Hubbard dimer, and Section~\ref{sec:num} explores them numerically for H$_4^{2+}$ and H$_2$. Section~\ref{sec:summary} concludes with a summary and outlook.

\section{Correlation measures}
\label{sec:corr-measures}

We denote by $\RDMset$ the set of all 1RDMs that arise from pure or ensemble $N$-particle states, and similarly by $\tRDMset$ the set of all 2RDMs~\cite{coleman2000-RDM-book}. For simplicity, we assume the one-particle Hilbert space $\mathcal{H}_1$ to be finite-dimensional, $d=\dim\mathcal{H}_1<\infty$, as is the case for a finite basis set or a discretization scheme.
Every $\g\in\RDMset$ is a Hermitian operator on $\mathcal{H}_1$ and thus has a spectral representation
\begin{equation}\label{1RDM}
\g = \sum_i n_i \ket{\chi_i}\bra{\chi_i}.
\end{equation}
The eigenvalues $\{n_i\}$, always listed in descending order, are the \emph{natural occupation numbers} (NONs) and the eigenvectors $\{\chi_i\}$ the \emph{natural orbitals} (NOs) of the underlying $N$-electron state. For fermions, the Pauli exclusion principle implies $0\leq n_i\leq 1$.

How can the correlation contained in an $N$-electron state be quantified? The most systematic answer to this question comes from quantum information theory, whose central aim is to characterize and quantify correlation and entanglement~\cite{Horodecki2009}. Its paradigm, based on local operations and classical communication (LOCC), rests on the notion of independent subsystems~\cite{Zanardi2001,Zanardi2002,Zanardi2004,Horodecki2009}. For $N$ electrons, this would require splitting the algebra of observables, e.g., as $\mathcal{A}_N\cong\mathcal{A}_1\otimes\mathcal{A}_{N-1}$, separating one electron from the remaining ones, which can be shown to be impossible for indistinguishable particles~\cite{AlivertiPiuri2024}. Electron correlation is thus not correlation between electrons in the strict quantum-information-theoretical sense, and therefore no unambiguous, universally accepted measure of it can be expected. Instead, electron correlation can only be quantified ad hoc, in a way that depends on the perspective or method at hand, be it based on wave functions, 2RDMs, or 1RDMs. All these perspectives share, however, a common notion of which pure states are uncorrelated: single Slater determinants, the eigenstates of non-interacting electron systems. Crucially, this notion can be expressed entirely in terms of the 1RDM. For $N$ particles in a single-determinant pure state, the NONs are
\begin{eqnarray}
    n_i = \left\{ \begin{array}{ll}
         1 & \text{for}\;1\leq i\leq N, \\
         0 & \text{for}\;i>N.
    \end{array}\right.
\end{eqnarray}
In other words, the 1RDM is idempotent, $\g^2=\g$, if and only if the state is pure and has the form of a single determinant (built from the NOs $\chi_i$)~\cite{Coleman1963}. Consequently, any pure state that is not a single determinant, and is thus correlated, has fractional NONs.

For ensemble states, the situation changes: fractional NONs can then also arise from classical mixing, for instance at finite temperature, so that non-idempotency alone no longer signals correlation. This simple observation is the key to Section~\ref{sec:corr-temp}, where correlation is modeled by an auxiliary thermal ensemble. It also suggests how to extend the notion of uncorrelated states: just as Slater determinants are the eigenstates of non-interacting systems, their grand-canonical thermal states are regarded as uncorrelated. These are the \emph{gauge-invariant quasi-free states} of algebraic quantum field theory~\cite[Ex.~5.2.20]{Brattelli-Robinson-book}, which in the literature on correlation measures are simply called \emph{free states}~\cite{Gottlieb2005,Gottlieb2007Properties,Gottlieb2017Nonfreeness,AlivertiPiuri2024}. Free states are grand-canonical states on Fock space for which the expectation value of \emph{every} many-body observable can be calculated exactly from the 1RDM, so that the full density matrix can be reconstructed from $\g$ alone~\cite{Cheong2004}; we denote by $\G_\g^\mathrm{free}$ the free state with 1RDM $\g$. Free states thus contain only one-body information. They include single Slater determinants as zero-temperature limits and are precisely the mean-field states referred to in the Introduction: by Wick's theorem, the 2RDM of $\G_\g^\mathrm{free}$ is exactly the mean-field expression $\Gmf(\g)$ introduced in Eq.~\eqref{eq:mf_2rdm} below~\cite{BachLiebSolovej1994}. Since the set $\RDMset$ of admissible 1RDMs is identical for canonical and grand-canonical ensembles~\cite{Coleman1963}, which would not be the case for $\tRDMset$, working on Fock space poses no restriction at the level of 1RDMs.

The correlation of an arbitrary state $\G$ can then be quantified by its distance from the set of free states, measured by the quantum relative entropy~\cite{Umegaki1962,Vedral2002}, the quantum analogue of the classical Kullback--Leibler divergence~\cite{Kullback-Leibler1951},
\begin{equation}
    D(\G_1\|\G_2) := \trace[\G_1(\log\G_1-\log\G_2)].
\end{equation}
Although the relative entropy is not a proper distance function, as it is neither symmetric nor does it obey the triangle inequality, it serves as a measure of distinguishability, since it is always nonnegative and vanishes if and only if $\G_1=\G_2$. Its minimum over all free states is attained for the free state with the same 1RDM and can be evaluated explicitly~\cite{GottliebMauser2014},
\begin{eqnarray}
    D(\G\|\G_{\g(\G)}^\mathrm{free}) &=& \min\{D(\G\|\G') : \G'\;\text{is free}\} \nonumber\\
    &=& -\sum_i (n_i\log n_i + (1-n_i)\log(1-n_i)) \nonumber \\
    && +\trace[\G\log\G].
\end{eqnarray}
This quantity is the \emph{nonfreeness} of $\G$, introduced by Gottlieb and Mauser~\cite{Gottlieb2005,Gottlieb2007Properties}. Here, the $\{n_i\}$ are the NONs of $\G$; the first term is the von Neumann entropy of the free state $\G_{\g(\G)}^\mathrm{free}$, and the second term is the negative von Neumann entropy of $\G$. This is a crucial observation: apart from the entropy of the state itself, the nonfreeness depends only on the 1RDM. For pure states, $\trace[\G\log\G]=0$, so the nonfreeness is determined by the NONs alone and equals the entropy of the closest free state. We thus arrive at a correlation measure for pure states that is a simple function of the 1RDM, the nonfreeness
\begin{equation}\label{eq:nf}
    \zeta_{\rm nf}(\g) = \Sph(\g) := -\sum_i (n_i\log n_i + (1-n_i)\log(1-n_i)),
\end{equation}
which we denote by $\Sph(\g)$ in later sections, reflecting its particle-hole symmetry. It coincides with the von Neumann entropy of an ideal Fermi gas with discrete modes and is directly related to the Fermi-Dirac distribution, as seen in Eq.~\eqref{eq:ni-dist}. The same function of the NONs has also appeared in other contexts, for instance as a correlation entropy~\cite{GoriGiorgi2002}, as the total orbital correlation minimized over all one-particle bases~\cite{GigenaRossignoli2015,Gigena2020,AlivertiPiuri2024}, and in empirical studies of electron interactions along reaction paths~\cite{FloresGallegos2023}.

The nonfreeness is, however, not the only conceivable correlation measure. To characterize suitable 1RDM-based measures more generally, we model a correlation measure as a function $\zeta:\RDMset\to\mathbb{R}_{\geq 0}$. Such a measure is meant to quantify the correlation of \emph{pure} $N$-electron states, such as ground states, whose fractional NONs reflect the electron-electron interaction. For mixed states, a function of the 1RDM alone cannot serve this purpose, since fractional NONs may then also stem from classical mixing; accordingly, the nonfreeness of a mixed state also involves the entropy of the state itself. We nevertheless define $\zeta$ on the convex set $\RDMset$ for technical convenience: unlike the set of pure-state representable 1RDMs~\cite{Altunbulak2008}, it is explicitly known and constitutes the natural domain of RDMFT~\cite{Valone80}. We now collect a set of basic properties that we require from $\zeta$. Some of these properties are necessary for any function $\zeta$ that measures correlation. Others ensure that the resulting method, which incorporates correlation effects through the energy and fractional occupation numbers, retains the computational cost of HF.
As shown in Section~\ref{sec:corr-temp}, this is guaranteed primarily if the measure does not involve the NOs at all. That the measure depends on the NONs alone can also be motivated by requiring invariance under orbital rotations. We formulate a total of six natural conditions.

\begin{enumerate}[leftmargin=*]
    \item\label{item:NON} The correlation measure depends \emph{only} (and explicitly) on the NONs $\{n_i\}$, so that $\zeta(\g)=\zeta(\{n_i\})$.
    \item\label{item:sum} The correlation measure has an additive form in terms of the NONs, without any cross terms, so that $\zeta(\g)=\sum_i f(n_i)$ with $f:[0,1]\to\mathbb{R}_{\geq 0}$ quantifying the correlation in each orbital.
    \item\label{item:idempotent} A single-determinant state carries no correlation, so $\zeta(\g)=0$ if and only if $\g^2=\g$. In this sense, every correlation measure with this property is also a measure of non-idempotency. Since an idempotent 1RDM has all NONs equal to either 0 or 1, this means $f(0)=f(1)=0$ and $f(n)>0$ for $0<n<1$.
    \item The correlation measure is invariant under the particle-hole transformation $n_i\mapsto 1-n_i$, i.e., $f(n)=f(1-n)$: exchanging the roles of particles and holes does not change the correlation. Importantly, this \emph{particle-hole symmetry} is also an exact condition for correlation functionals in RDMFT~\cite{Yasuda2001}.
    \item The correlation measure is \emph{Schur-concave}, i.e., monotonic with respect to the majorization order of the NONs~\cite{Marshall2011-book}\footnote{For two vectors $\mathbf{x},\mathbf{y}\in\mathbb{R}^d$ with entries sorted in descending order, $x_1^\downarrow\geq\dots\geq x_d^\downarrow$, one says that $\mathbf{x}$ \emph{majorizes} $\mathbf{y}$, written $\mathbf{x}\succ\mathbf{y}$, if $\sum_{i=1}^k x_i^\downarrow\geq\sum_{i=1}^k y_i^\downarrow$ for all $k=1,\dots,d-1$, with equality for $k=d$~\cite{Marshall2011-book,Bhatia1997}. Equivalently, $\mathbf{y}=D\mathbf{x}$ for some doubly stochastic matrix $D$, so that $\mathbf{y}$ is a more disordered, `mixed-up' version of $\mathbf{x}$. Majorization is thus the standard way of comparing the disorder of probability-like vectors and underlies the theory of entanglement transformations~\cite{Nielsen1999}. For Hermitian matrices, majorization is defined through their spectra; for 1RDMs, $\g_1\succ\g_2$ means that the NONs of $\g_1$ majorize those of $\g_2$. A function $\zeta$ is Schur-concave if $\mathbf{x}\succ\mathbf{y}$ implies $\zeta(\mathbf{x})\leq\zeta(\mathbf{y})$. Since all admissible NON vectors satisfy $0\leq n_i\leq 1$ and $\sum_i n_i=N$, the NONs $(1,\dots,1,0,\dots,0)$ of a single Slater determinant majorize all of them, whereas the uniform NONs $n_i=N/d$ are majorized by all of them.}: if the NONs of one state majorize those of another, the former is at most as correlated as the latter. Since the NONs of a single Slater determinant majorize all other admissible NONs, this condition orders states from uncorrelated towards increasingly spread-out occupations. It parallels pure-state entanglement theory, where entanglement monotones are precisely the Schur-concave functions of the Schmidt coefficients~\cite{Nielsen1999,Vidal2000}, and majorization relations of the 1RDM spectrum similarly govern one-body entanglement in fermionic systems~\cite{Gigena2020}. For the additive form of condition~\ref{item:sum}, Schur-concavity is equivalent to \emph{concavity} of $f(n)$~\cite{Marshall2011-book}, which in turn renders $\zeta(\g)$ concave on $\RDMset$.
    \item\label{item:strictly-bijective} In order to uniquely map the orbital energies of a mean-field approximation to NONs (as in Eq.~\eqref{eq:LM-ni} below), $f(n)$ must be \emph{differentiable} on $n\in(0,1)$ with a \emph{bijective} derivative that maps $(0,1)\to\mathbb{R}$. This also renders $\zeta(\{n_i\})$ \emph{strictly} Schur-concave.
\end{enumerate}

Condition~\ref{item:strictly-bijective} also appears in \citet{martinez2023} for a given distribution of NONs. It is straightforward to verify that the nonfreeness $\Sph$ satisfies conditions~\ref{item:NON}--\ref{item:strictly-bijective}. The connection between correlation measures and the correlation functional will be established in Section~\ref{sec:corr-temp}. First, however, we discuss further measures.

A second correlation measure, with a square-root (semicircular) profile, is defined as
\begin{equation}\label{eq:zeta-sqrt}
    \zeta_{\rm sqrt}(\g):=\frac 1 2 \sum_i \sqrt{n_i(1-n_i)}.
\end{equation}
In Section~\ref{sec:Hubbard}, this form will be shown to be proportional to the \emph{exact} RDMFT correlation functional of the symmetric Hubbard dimer. It also satisfies all conditions and, in addition, reproduces the known boundary behavior of the exact correlation functional. Since by condition~\ref{item:strictly-bijective} the derivative of $f(n)$ must cover all of $\mathbb{R}$, the derivative $\partial\zeta/\partial n_i$ necessarily diverges at the boundary of the interval $[0,1]$. For the correlation functional of RDMFT, this phenomenon was previously described as the `exchange force'~\cite{schilling2019b}, with the functional varying to lowest order as $\sqrt{n}$ close to the boundary, precisely as in Eq.~\eqref{eq:zeta-sqrt}.
The same phenomenon appears generally in functional theories~\cite{Maciazek2021,Wang2026}, so it could be demanded as a refined version of condition~\ref{item:strictly-bijective}. A consequence thereof will be demonstrated numerically in Section~\ref{sec:combining-measures}.

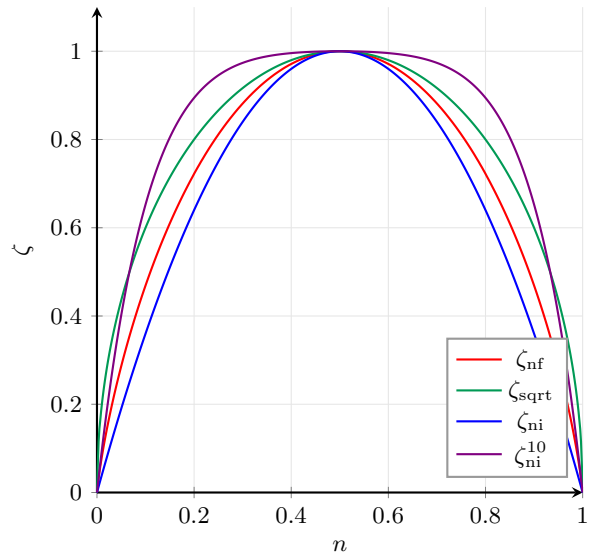
\begin{figure}[ht]
\begin{tikzpicture}
\begin{axis}[
    width=8cm, height=8cm,
    xlabel={$n$},
    ylabel={$\zeta$},
    xmin=0, xmax=1,
    ymin=0, ymax=1.1,
    domain=0:1,
    samples=300,
    smooth,
    axis lines=left,
    grid=major,
    grid style={gray!20},
    legend pos=south east,
    legend style={
        draw=black!40,
        fill=white,
        font=\small
    },
    tick label style={font=\small},
    label style={font=\small},
    thick
]

\addplot[
    domain=0.0001:0.9999,
    samples=300,
    color=red,
    thick
] {-(x*ln(x) + (1-x)*ln(1-x)) / ln(2)};
\addlegendentry{$\zeta_\mathrm{nf}$}

\addplot[
    domain=0:1,
    samples=300,
    color=ForestGreen,
    thick
] {1/0.5*sqrt(x*(1-x))};
\addlegendentry{$\zeta_\mathrm{sqrt}$}

\addplot[
    domain=0:1,
    samples=300,
    color=blue,
    thick
] {1/0.5^2*x*(1-x)};
\addlegendentry{$\zeta_\mathrm{ni}$}

\addplot[
    domain=0:1,
    samples=300,
    color=violet,
    thick
] {((x-x^10)+(1-x-(1-x)^10)) / 0.998047};
\addlegendentry{$\zeta_\mathrm{ni}^{10}$}

\end{axis}
\end{tikzpicture}
\caption{Single-orbital contributions $f(n)$ of the correlation measures discussed in the text, each rescaled to a maximum value of 1 at $n=1/2$.}
\label{fig:corr-measures}
\end{figure}

We also mention the non-idempotency measure
\begin{eqnarray}\label{eq:zeta_non_idem}
    \zeta_{\rm ni}(\g) := \sum_i n_i(1-n_i)
\end{eqnarray}
with a parabolic profile, which is the standard measure for probing the idempotency $\g=\g^2$ of the 1RDM. The factor $\frac 1 2$ in Eq.~\eqref{eq:zeta-sqrt} was chosen such that the maximal values of $\zeta_{\rm sqrt}$ and $\zeta_{\rm ni}$ agree. The non-idempotency measure was suggested as an RDMFT functional to augment DFT methods by energetically favoring non-idempotent 1RDMs~\cite[Eq.~(5)]{Gibney2022}. It was also used as a descriptor of non-dynamic correlation, while $\zeta_{\rm sqrt}(\g)$ represents the full correlation, so that $\zeta_{\rm sqrt}(\g)-\zeta_{\rm ni}(\g)$ roughly quantifies the dynamic part~\cite{RamosCordoba2016,RamosCordoba2017}.
In the asymmetric Hubbard dimer, $\zeta_{\rm ni}$ appears as an estimate for the correlation functional in Eqs.~\eqref{eq:Hubbard-Ecum-estimates} and \eqref{eq:hub_dim_corr_func_approx}.
Note, however, that $\zeta_{\rm ni}$ violates condition~\ref{item:strictly-bijective}, since its derivative remains bounded; as a consequence, the corresponding minimizers can become idempotent, as observed in Section~\ref{sec:Hubbard}. The generalized form (here in its symmetrized variant)
\begin{eqnarray}\label{eq:zeta_non_idem_q}
    \zeta_{\rm ni}^q(\g) := \sum_i((n_i- n_i^q) + (1-n_i- (1-n_i)^q))
\end{eqnarray}
of the non-idempotency measure has also been considered for quantifying correlation~\cite{Ziesche1995}. It is related to the entropy $S^q(\g)=\zeta_{\rm ni}^q(\g)/(q-1)$ introduced by \citet{Tsallis1988} for the statistics of nonextensive systems. For $q=2$, Eq.~\eqref{eq:zeta_non_idem} is recovered up to a factor of 2, while the entropy $S^q(\g)$ reduces to $\Sph(\g)$ in the limit $q\to 1$. The R\'enyi entropies form another family with the same limit. A graphical comparison of the correlation measures discussed above is given in Fig.~\ref{fig:corr-measures}.

\section{Correlation temperature}
\label{sec:corr-temp}

Having established correlation measures based on the 1RDM, we now turn them into a computational method. Following the guiding idea of the Introduction, we model the correlation energy as a function of a correlation measure $\zeta(\g)$. Since the measures of Section~\ref{sec:corr-measures} depend only on the 1RDM, this naturally places the construction within RDMFT~\cite{UnifiedFT2026-preprint}: in a coarse-graining step, the correlation functional, which in general depends on the full 1RDM, is replaced by a function of the single scalar variable $\zeta$. This step goes along with the introduction of a scope, i.e., the set of Hamiltonians that model the quantum-chemical systems of interest. We develop the construction for a general measure satisfying the conditions of Section~\ref{sec:corr-measures} and specialize to the nonfreeness $\Sph(\g)$ of Eq.~\eqref{eq:nf} where this leads to particularly transparent results.

RDMFT addresses the ground-state problem for Hamiltonians $\hat H = \hat h + \hat W$ of an $N$-particle system, with $\hat h$ the one-body contribution, typically comprising the kinetic energy and the external potential, and $\hat W$ the two-body interaction. Since the interaction $\hat W$ is fixed, while $\hat h$ can be any Hermitian one-particle operator and couples to the state only through $\trace[\hat h\g]$, the 1RDM $\g$ is singled out as the natural reduced variable replacing the full quantum state~\cite{liebert23,UnifiedFT2026-preprint}.
Within this setting, the universal functional $\F(\g)$ is defined by a constrained search over all $N$-representable 2RDMs $\G^{(2)}\in\tRDMset$ that lead to a given $\g\in\RDMset$ (written as ``$\G^{(2)} \mapsto \g$''),
\begin{equation}\label{eq:F-def}
    \F(\g) := \inf_{\G^{(2)} \mapsto \g} \trace[\hat W \G^{(2)}].
\end{equation}
The ground-state energy then follows from a variation over all admissible 1RDMs instead of the full quantum state,
\begin{equation}\label{eq:E0-variation}
    E_0 = \inf_\g\left\{ \trace[\hat h \g] + \F(\g) \right\}.
\end{equation}
A first, obvious approximation to $\F(\g)$ is obtained from a mean-field approximation to the 2RDM in terms of the 1RDM, where we use the expansion in terms of NOs from Eq.~\eqref{1RDM},
\begin{eqnarray}
  \label{eq:mf_2rdm}\Gmf(\g) &=& \frac{1}{2}(\hat I - \hat X)\,\g \otimes \g, \\
  \label{eq:mfFunc}
  \Fmf(\g) &=& \trace[\hat W \Gmf(\g)] \nonumber \\
           &=& \frac{1}{2} \sum_{i,j} n_i n_j \bra{\chi_i \chi_j} \hat W \ket{\chi_i \chi_j}.
\end{eqnarray}
Here, the exchange operator is defined as $\hat X(\ket{\chi_i}\otimes\ket{\chi_j})=\ket{\chi_j}\otimes\ket{\chi_i}$, and $\ket{\chi_i \chi_j} = (\ket{\chi_i}\otimes\ket{\chi_j} - \ket{\chi_j}\otimes\ket{\chi_i})/\sqrt{2}$ is a two-orbital Slater determinant.
For a general 1RDM $\g$, the expression $\Gmf(\g)$ is \emph{not} the 2RDM of any $N$-particle state, pure or mixed, and does not contract back to $\g$ in the usual way. Rather, as noted in Section~\ref{sec:corr-measures}, it is the 2RDM of the grand-canonical free state $\G_\g^\mathrm{free}$ on Fock space. Only if $\g$ is idempotent, and thus represents a single-determinant pure state, is $\Gmf(\g)$ precisely the corresponding $N$-particle 2RDM. In this way, the functional $\Fmf(\g)$ is connected to the HF method built on Slater determinants~\cite{LiebHF, Lieb1977}, and it serves as the basis for more refined approximations within RDMFT. To this end, we define the correlation functional as the difference between the exact functional and $\Fmf(\g)$,
\begin{equation}\label{eq:Fc}
    \Fc(\g) := \F(\g) - \Fmf(\g) \leq 0.
\end{equation}
For this inequality to hold for all $\g\in\RDMset$, the interaction $\hat W$ must be positive semidefinite~\cite{LiebHF,Levy87}, as is the case for the most relevant interactions, such as the Coulomb interaction in quantum chemistry, the on-site interaction of the Hubbard model, and the repulsive contact interaction in ultracold atomic gases. We stress that at this level the decomposition $\F(\g)=\Fmf(\g)+\Fc(\g)$ is still \emph{exact}, in the sense that the variation in Eq.~\eqref{eq:E0-variation} yields the exact ground-state energy and 1RDM.

Building on this exact decomposition, the central idea is to coarse-grain the correlation functional: rather than depending on the full 1RDM, it is approximated as a function of a single correlation measure,
\begin{equation}\label{eq:F-tilde}
    \tilde\F(\g)=\Fmf(\g)+\tilde\Fc(\zeta(\g)).
\end{equation}
How should the function $\tilde\Fc$ be chosen? A natural requirement is that it never overestimates the magnitude of the correlation energy, $\tilde\Fc(\zeta(\g))\geq\Fc(\g)$, so that $\tilde\F$ is bounded from below by the exact functional. The sharpest function satisfying this requirement on a given set of 1RDMs is obtained by taking, for each value of $\zeta$, the supremum of $\Fc$ over all 1RDMs sharing this value. Taken over all 1RDMs, however, such a supremum can be very conservative, since a single value of $\zeta$ is shared by 1RDMs of very different systems. Yet in many situations only a limited set of systems needs to be considered, e.g., the dissociation of a certain molecule. We therefore restrict the supremum to the set $\mathcal{G}(\mathcal{S})$ of ground-state 1RDMs (including those of ensemble ground states in the case of degeneracies) of a chosen \emph{scope} $\mathcal{S}$ of Hamiltonians of interest, such as various molecular geometries~\cite{UnifiedFT2026-preprint}, and define the coarse-grained correlation functional as
\begin{equation}\label{eq:coarse-func}
    \tilde\Fc(\zeta) = \sup_{\mathcal{G}(\mathcal{S})\ni\g\mapsto\zeta}\Fc(\g),
\end{equation}
where the supremum is taken over all $\g\in\mathcal{G}(\mathcal{S})$ with $\zeta(\g)=\zeta$, and $\tilde\Fc(\zeta)$ is set to zero if $\zeta$ is not attained in $\mathcal{G}(\mathcal{S})$. By construction, $\tilde\Fc(\zeta)\leq 0$ and $\tilde\Fc(\zeta(\g))\geq\Fc(\g)$ for all $\g\in\mathcal{G}(\mathcal{S})$, so that
\begin{equation}\label{eq:Fc-inequ}
    \F(\g)\leq\tilde\F(\g)\leq\Fmf(\g)\quad\text{for all }\g\in\mathcal{G}(\mathcal{S}).
\end{equation}
If the variation is restricted to $\mathcal{G}(\mathcal{S})$, the resulting energy can thus never fall below the exact ground-state energy, and the narrower the scope, the sharper the coarse-grained functional, as illustrated for the Hubbard dimer in Section~\ref{sec:Hubbard}. In practice, $\mathcal{G}(\mathcal{S})$ is rarely known explicitly and the variation is carried out over the full set $\RDMset$, so that these guarantees no longer hold strictly; approximations such as the linear form discussed below are then obtained by fitting rather than from the supremum.

The approximate variational problem for the ground state then reads
\begin{equation}\label{eq:E0-variation-approx}
\begin{aligned}
    \tilde E_0 &= \inf_\g\left\{ \trace[\hat h \g] + \tilde\F(\g) \right\} \\
    &= \inf_\g\left\{ \trace[\hat h \g] + \Fmf(\g) + \tilde\Fc(\zeta(\g)) \right\}.
\end{aligned}
\end{equation}
When varying over $\g\in\RDMset$ in terms of NOs and NONs with the usual Lagrange-multiplier approach, the term $\tilde\Fc(\zeta(\g))$ drops out of the variation with respect to the NOs, as it depends on the NONs only. Because $\Fmf(\g)$ corresponds exactly to the HF functional with fractional occupation numbers, this yields the usual HF equations, with the only difference that the orbital-dependent operators are weighted by the occupation numbers $n_j$,
\begin{equation}\label{eq:th-HF}
    \left(\hat h + \sum_j n_j(\hat J_j - \hat K_j)\right) \chi_i = \varepsilon_i \chi_i.
\end{equation}
Here, $\hat J_j$ and $\hat K_j$ are the usual Hartree and exchange operators of HF theory. The variation with respect to the NONs must take $\tilde\Fc(\zeta)$ into account. For a measure of the additive form $\zeta(\g)=\sum_i f(n_i)$, the chain rule yields (assuming differentiability)
\begin{equation}\label{eq:LM-ni}
    \varepsilon_i + \frac{\partial \tilde\Fc}{\partial \zeta}\, f'(n_i) = \mu,
\end{equation}
where the Lagrange multiplier $\mu$, the chemical potential, is fixed by the normalization $\sum_in_i=N$. Defining the variable conjugate to the measure,
\begin{equation}
    \kc(\zeta):=-\frac{\partial \tilde\Fc(\zeta)}{\partial \zeta},
\end{equation}
assuming $\kc>0$, and using that $f'$ is a bijection from $(0,1)$ to $\mathbb{R}$ (condition~\ref{item:strictly-bijective} of Section~\ref{sec:corr-measures}), Eq.~\eqref{eq:LM-ni} is uniquely solved by
\begin{equation}\label{eq:ni-general}
    n_i = (f')^{-1}\!\left(\frac{\varepsilon_i - \mu}{\kc(\zeta)}\right).
\end{equation}
The variable $\kc$ thus sets the energy scale on which the occupation numbers are distributed over the orbital energies, much like a temperature. For the nonfreeness, $f(n)=-n\log n-(1-n)\log(1-n)$ and hence $f'(n)=\log(1-n)-\log n$, so that Eq.~\eqref{eq:ni-general} becomes the Fermi-Dirac distribution
\begin{equation}\label{eq:ni-dist}
    n_i = \left(1 + \rme^{(\varepsilon_i - \mu)/\kc(\zeta)}\right)^{-1}.
\end{equation}
Note that $\kc$ still depends on all $\{n_i\}$ through $\zeta$, but Eqs.~\eqref{eq:th-HF} and \eqref{eq:ni-general} must anyway be solved self-consistently. In view of Eq.~\eqref{eq:ni-dist}, we call $\kc$ the \emph{correlation temperature}. It is an auxiliary quantity: an effective temperature that serves as a surrogate for correlation effects. As noted in the Introduction, several methods in the quantum-chemical literature build on a similar idea. Here, however, the fractional occupations follow directly from quantifying electron correlation, which distinguishes them from an effective temperature introduced for purely practical reasons, namely to help with $v$-representability~\cite{Ullrich2002,Kraisler2009,Woods2019}, to guarantee numerical stability~\cite{Weinert1992,Wentzcovitch1992,Marzari1997,Rabuck1999}, or to overcome the so-called ``correlation paradox''~\cite{Ding2020}. A correlation temperature can nevertheless be expected to share these beneficial side effects.

For the nonfreeness and a fixed value of $\kc$, the coupled equations~\eqref{eq:th-HF} and \eqref{eq:ni-dist} correspond exactly to the thermal HF method in the grand-canonical ensemble~\cite{Mermin1963}. Indeed, since the nonfreeness of a pure state equals the entropy of the closest free state, the functional minimized in Eq.~\eqref{eq:E0-variation-approx} with a linear dependence $-\kc\Sph(\g)$ is, up to a constant, precisely the grand-canonical free energy of thermal HF at temperature $\kc$. In particular, the method can be executed at mean-field cost. Only in the grand-canonical ensemble, which allows for particle-number fluctuations around $N$, does a Fermi-Dirac distribution arise from a thermal Gibbs state. This may seem surprising at first, since our description concerns a molecular system with fixed particle number at zero temperature. The open system, however, serves merely as an auxiliary reference system that shares important physical properties, and only through it can the notions of temperature and chemical potential be sensibly introduced~\cite{Baerends2022}. The Fermi-Dirac form of Eq.~\eqref{eq:ni-dist} is thus specific to the nonfreeness; other correlation measures lead to other distributions via Eq.~\eqref{eq:ni-general}.

A practical implementation relies on a reliable approximation to the coarse-grained correlation functional $\tilde\Fc(\zeta)$, or equivalently to the correlation temperature $\kc(\zeta)$, for as broad a range of systems as possible. Since the interaction is fixed in the RDMFT functional $\F(\g)$, the variable parts of the theory are the particle number and the one-particle Hamiltonian $\hat h$, including the external potential. In this respect, the correlation temperature $\kc(\zeta)$ plays a role analogous to the exchange-correlation potential of Kohn-Sham DFT~\cite{KS1965}. The ground state is then approximated by a self-consistent solution of Eq.~\eqref{eq:th-HF}, with the $\{n_i\}$ determined self-consistently from Eq.~\eqref{eq:ni-general}, where $\kc(\zeta)$ in turn depends on the occupation numbers.
Once converged, with $\g_*$ the resulting 1RDM and $\zeta_*=\zeta(\g_*)$, the approximate ground-state energy is evaluated as
\begin{equation}
\begin{aligned}
    \tilde E_0 &= \trace[\hat h\g_*] + \Fmf(\g_*) + \tilde\Fc(\zeta_*).
\end{aligned}
\end{equation}

In the spirit of Jacob's ladder~\cite{Perdew2001}, possible approximations can be organized into rungs. On the zeroth rung, the coarse-grained correlation functional $\tilde\Fc$ is simply a constant that depends only on certain chemical information. This implies a vanishing correlation temperature, $\kc=0$, so that the method reduces to a standard HF calculation followed by a correlation-energy correction. This is essentially the idea of the ``ultrafast correlation energy estimator'', which adds a fixed partial correlation energy for each bond type~\cite{Witkowski2025} and is called `ultrafast' because it can be applied on top of a HF calculation at virtually no additional cost. The same basic idea underlies ``bond additivity corrections''~\cite{Melius2000}.
Any more involved coarse-grained correlation functional also keeps the cost at the mean-field level, requiring only a thermal-HF-like self-consistent calculation with nonzero temperature.
On the first rung of the ladder, the coarse-grained functional genuinely depends on the correlation measure, which thereby enters the self-consistent procedure.
In the simplest case, this is a linear approximation of the form
\begin{equation}\label{eq:tildeFc-linear}
    \tilde\Fc(\zeta) \approx -\kc\zeta - b,
\end{equation}
which corresponds to a constant correlation temperature and can serve as an efficient approximation for $\Fc(\g)$, although it is not guaranteed to respect the bounds of Eq.~\eqref{eq:Fc-inequ} in every setting.
It was reported that, for a scope restricted to diatomic dissociation processes, $\Fc(\g)$ indeed approximately follows this linear behavior for the nonfreeness~\cite{WangKnowles21-PRA}, in line with a ``correlation conjecture''~\cite{Graf2026} that is closely related to the ``Collins conjecture''~\cite{Collins}. Note that this observation alone does not confirm the validity of Eq.~\eqref{eq:tildeFc-linear}, since it only requires $\Fc(\g)\approx -\kc \Sph(\g) - b$ for the ground-state 1RDMs $\g$ along the dissociation curve.
The observed linear relation between $\Sph(\g)$ and $\Fc(\g)$ led to the \iDMFT{} method~\cite{WangBaerends22-PRL}, which employs Eq.~\eqref{eq:tildeFc-linear} with fixed $\kc$ and $b$ and which we consequently place on the first rung of the ladder. Its parameters are usually optimized directly towards accurate energies~\cite{Irimia2023-open-shell,Irimia2023-correlation}.
Functional approximations including terms of second order in $\zeta$ have also been considered and tested numerically~\cite{Hu2024-dispersion}.
Higher rungs could use additional information extracted from the 1RDM. Our derivation suggests that such developments should focus on a more general correlation temperature $\kc(\zeta, [\g])$ that depends on the measure and, potentially, on further information contained in the 1RDM. Self-consistent schemes for determining effective electronic temperatures are known from TAO-DFT~\cite{Lin2017,Chen2022}, and \citet{Gibney2023} derived an estimate for $\kc$ from electron repulsion integrals, coupled to the non-idempotency measure $\zeta_{\rm ni}$.

\section{Illustration with the Hubbard dimer}
\label{sec:Hubbard}

The concepts from the previous section will be demonstrated here using the simple, explicit example of an (asymmetric) Hubbard dimer at half-filling. Interestingly, this setting naturally leads to an occurrence of two of the considered correlation measures, concretely $\zeta_\mathrm{ni}$ and $\zeta_\mathrm{sqrt}$.
We consider the Hamiltonian
\begin{equation}\label{eq:hub_ham}
    \begin{split}
        &\left.
        \begin{aligned}
        \hat{H}_\mathrm{HD}=
        &-\sum_{\sigma\in\{\uparrow,\downarrow\}}\Big(t\hspace{2pt}\hat{f}^\dagger_{1\sigma}\hat{f}_{2\sigma}+\mathrm{h.c.}\Big)\\
        &+\frac{V}{2}\sum_{\sigma\in\{\uparrow,\downarrow\}}\Big(\hat{f}^\dagger_{1\sigma}\hat{f}_{1\sigma}-\hat{f}^\dagger_{2\sigma}\hat{f}_{2\sigma}\Big)
        \end{aligned}\quad\right\}=:\hat{h}_{t,V}
        \\
        &\left.\hspace{33.4pt}+\;U\sum_{j\in\{1,2\}}\hat{f}^\dagger_{j\uparrow}\hat{f}_{j\downarrow}^\dagger\hat{f}_{j\downarrow}\hat{f}_{j\uparrow}. \hspace{39pt}\right\}=:\hat{W}
    \end{split}
\end{equation}
The operator $\hat{f}^{(\dagger)}_{i\sigma}$ denotes the canonical fermionic annihilation (creation) operator for a particle at site $i$ with spin polarization $\sigma$. For our example, we consider a real hopping parameter $t\in\mathbb{R}$, a difference in local potentials $V\in\mathbb{R}$, and a positive on-site interaction at strength $U\in\mathbb{R^+}$. The single-particle terms of Eq.~\eqref{eq:hub_ham} are collected in the one-particle Hamiltonian $\hat{h}_{t,V}$, while the pair-interaction is again denoted by $\hat{W}$.
Through its high versatility, combined with its analytical solvability, the Hubbard dimer has assumed a role as a paradigmatic test system for the study of strong correlation and functional approximations~\cite{carrascal2015a}.

For the ground state search of the Hamiltonian in Eq.~\eqref{eq:hub_ham}, we consider states supported on the singlet sector of the two-particle Hilbert space. This subspace is spanned by
\begin{equation}\label{eq:hub_dim_singlet_onb}
    \begin{split}
        &\ket{\Phi_1}=\hat{f}^\dagger_{1\uparrow}\hat{f}^\dagger_{1\downarrow}\ket{0},\\
        &\ket{\Phi_2}= \hat{f}^\dagger_{2\uparrow}\hat{f}^\dagger_{2\downarrow}\ket{0},\\
        &\ket{\Phi_3}=\frac{1}{\sqrt{2}}\Big(\hat{f}^\dagger_{1\uparrow}\hat{f}^\dagger_{2\downarrow}-\hat{f}^\dagger_{1\downarrow}\hat{f}^\dagger_{2\uparrow}\Big)\ket{0},
    \end{split}
\end{equation}
where $\ket{0}$ denotes the vacuum state.

In order to apply the framework from Sec.~\ref{sec:corr-temp}, a sensible choice for a correlation measure (in the sense of Sec.~\ref{sec:corr-measures}) must be identified. For this, we examine properties of the cumulant energy of ground states in our dimer system. Building on the definition of the two-body correlation matrix in Eq.~\eqref{eq:mf_2rdm}, the cumulant energy is defined as~\cite{kutzelnigg2003, Levy87}
\begin{equation}\label{eq:def_e_cum}
    \Ecum(\G)=\trace\left[\hat W\Gamma\right]-\trace\left[\hat{W}\Gmf(\g)\right],
\end{equation}
where $\G$ denotes an $N$-particle state, with $\g$ the corresponding 1RDM. The cumulant energy can be understood as the analogue of the correlation functional at the level of the full state; applying the constrained-search formalism yields the correlation functional, $\Fc(\g)=\inf_{\G\mapsto\g}\Ecum(\G)$.

Besides choosing a sensible correlation measure for the coarse-grained variable, a crucial step in our approximation procedure is specifying a \emph{scope} for our new functional. In prior work by our group, it was established that different scopes for the Hubbard dimer yield distinct sets of $v$-representable 1RDMs~\cite{liebert23}. For illustrating the procedure, the following two scopes are suited particularly well:
\begin{alignat}{3}\label{eq:hub_dim_scopes_1}
        &i)\qquad &\mathcal{S}_\mathrm{HD}^\mathbb{R} &:=\left\{\hat h_{t,V}+\hat W: t,V\in\mathbb{R}\right\},\\
        \label{eq:hub_dim_scopes_2}
        &ii) &\mathcal{S}_\mathrm{HD}^\mathrm{sym} &:=\left\{\hat h_{t,V}+\hat W: t\in\mathbb{R}, V=0\right\}.
\end{alignat}
The first scope includes the most general form of real-valued symmetric single-particle operators, while the second case corresponds to the setting of the (left-right) symmetric Hubbard dimer with hopping between sites.
Evidently, $\mathcal{S}_\mathrm{HD}^\mathrm{sym}\subset\mathcal{S}_\mathrm{HD}^{\mathbb{R}}$. 
Crucially, this allows for the derivation of a similar hierarchy for the sets of corresponding $v$-representable 1RDMs: $\mathcal{G}(\mathcal{S}_\mathrm{HD}^\mathrm{sym})\subseteq\mathcal{G}(\mathcal{S}_\mathrm{HD}^{\mathbb{R}})$.
For illustrative purposes, we restrict ourselves in this section to the sets of \emph{pure-state} $v$-representable 1RDMs, which simplifies the derivation of analytical formulas. From here on, we will denote these sets by $\tilde{\mathcal{G}}(\mathcal{S}_\mathrm{HD}^\mathrm{sym})$ and $\tilde{\mathcal{G}}(\mathcal{S}_\mathrm{HD}^{\mathbb{R}})$ to emphasize the restriction compared to the general set of ensemble $v$-representable 1RDMs.
The supremum applied during coarse-graining in Eq.~\eqref{eq:coarse-func} translates the relationship between sets into an inequality on the level of the correlation functional:
\begin{equation}\label{eq:hub_dim_ineq_scopes_func}
    \tilde{\mathcal{F}}^c_{\mathcal{S}_\mathrm{HD}^\mathrm{sym}}(\gamma)
    \leq\tilde{\mathcal{F}}^c_{\mathcal{S}_\mathrm{HD}^{\mathbb{R}}}(\gamma),
\end{equation}
where the subscript indicates the scope defining the domain of 1RDMs used for coarse-graining. Inevitably, the dependence on the domain also reflects on the validity range of the above inequalities, such that Eq.~\eqref{eq:hub_dim_ineq_scopes_func} holds for $\g\in\tilde{\mathcal{G}}(\mathcal{S}_\mathrm{HD}^\mathrm{sym})$.

From the coarse-graining, it is known that all of the above approximations are lower-bounded by the true correlation functional $\Fc$. Hence, Eq.~\eqref{eq:hub_dim_ineq_scopes_func} demonstrates how increased knowledge about inherent structure in the Hamiltonian holds the potential to improve the functional approximation directly via the coarse-graining procedure.

It shall also be pointed out that the scopes of Eqs.~\eqref{eq:hub_dim_scopes_1}~\&~\eqref{eq:hub_dim_scopes_2} in the Hubbard dimer capture physically highly relevant settings:
In particular, for real-valued hopping strengths, the Hubbard dimer can be interpreted as a minimalistic model of single-bond, diatomic molecules. 
In this picture, the parameter $t$ can be related in lowest order to the internuclear distance as $|t|\sim \exp(-\rho R)$~\cite{giesbertz2018}, where $\rho$ denotes the reciprocal characteristic length scale of the system. Hence, the reduced internuclear distance is related to the hopping as $\rho R\sim \log(1/|t|)$. The parameter $V$ quantifies the difference in (effective) nuclear charge between the nuclei.
From the particle-hole symmetry of the half-filled system, it follows that the physics between $\pm t$ are identical. The same holds true for the sign of the potential difference $V$ due to the left-right symmetry of the system.

\begin{figure}
    \centering
    \includegraphics[width=\columnwidth]{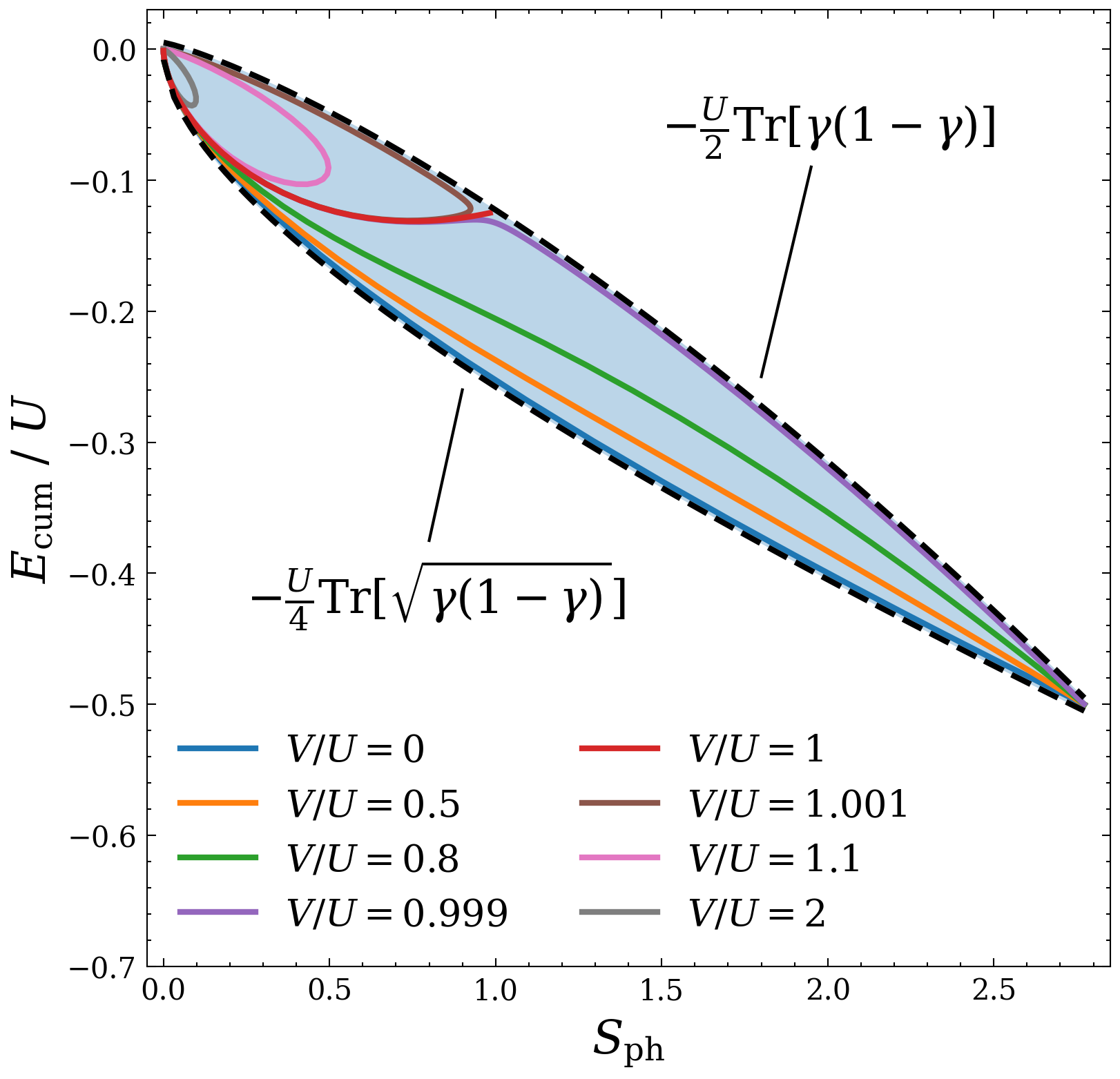}
    \caption{Hubbard dimer. Relationship between cumulant energy $\Ecum$ and ph entropy $\Sph$ during variation of the hopping strength $t\in\mathbb{R}^+_0$ for fixed values of $V$. The set of all tuples $(\Sph, \Ecum)$ accessible through ground states of $\mathcal{S}_\mathrm{HD}^{\mathbb{R}}$ is indicated by the blue shade. For the symmetric scope $\mathcal{S}_\mathrm{HD}^\mathrm{sym}$, the full set of $v$-representable tuples corresponds to the one-dimensional path corresponding to $V=0$.
    The interaction strength is fixed to $U=1$.}
    \label{fig:hub_dim_E_cum_S}
\end{figure}

Figure~\ref{fig:hub_dim_E_cum_S} illustrates the relationship between the two scopes introduced in Eqs.~\eqref{eq:hub_dim_scopes_1}~\&~\eqref{eq:hub_dim_scopes_2} on the level of cumulant energy $\Ecum$ plotted against a correlation measure complying with the constraints from Sec.~\ref{sec:corr-measures}. The ideal correlation measure is not known a priori. In the figure, we chose to default to the particle-hole symmetrized entropy $\Sph$, which has been shown to be in a linear relationship with the ground-state cumulant energy for dissociations in certain molecules~\cite{WangKnowles21-PRA,Graf2026}.
For fixed values of the potential difference $V$, the relationship is shown throughout the variation of $t$. Interpreting our Hubbard system as a minimal model for diatomic molecules, the variation of $t$ is analogous to nuclear dissociation. If $V<U=1$, it is found that all paths $(\Sph(t), \Ecum(t))$ run from $(4\log(2), -U/2)$ at $t=0$ to a fixed end point $(0,0)$ for $t\to\infty$. In the regime of $V>U$, the points for $t$ in both limits coincide at $(0,0)$. Consequently, for $V>U$, the paths of $(\Sph(t), \Ecum(t))$ for a variation of $t$ form closed loops. In the special case of $V=U$, the path, again, approaches $(0,0)$ for $t\to\infty$. However, in the limit $t\to0$, the path runs towards a unique point with $\Sph=8\log(2)-2\sqrt{3}\log\big(2+\sqrt{3}\big)\approx0.9831$ and $\Ecum=-U/8$.

The area covered by the set of all possible paths $(\Sph(t),\Ecum(t))$ is shaded in blue in Figure~\ref{fig:hub_dim_E_cum_S}. This corresponds to the full set of tuples $(\Sph,\Ecum)$ that are $v$-representable via Hamiltonians in $\mathcal{S}_\mathrm{HD}^\mathbb{R}$. For the symmetric scope $\mathcal{S}_\mathrm{HD}^\mathrm{sym}$, which corresponds to the case $V=0$, we find that the set of $v$-representable entropy-energy pairs is compressed into a one-dimensional line, agreeing with the lower bound of the shaded area.
The exact upper and lower bounds, as well as the functional approximations obtained for the two scopes under consideration, are discussed in the following Secs.~\ref{ssec:hub_dim_scope_real}~\&~\ref{sec:hub_dim_symmetric}.

\subsection{Full real-valued scope \texorpdfstring{$\mathcal{S}_\mathrm{HD}^\mathbb{R}$}{SHDR}}\label{ssec:hub_dim_scope_real}

Building on the definition in Eq.~\eqref{eq:def_e_cum}, it can be shown that for any (singlet) ground state $\G$ of a Hamiltonian in $\mathcal{S}_\mathrm{HD}^{\mathbb{R}}$, it holds that
\begin{equation}\label{eq:hub_e_cum_bound}
    -\frac{U}{2}\trace\big[\g(1-\g)\big]\geq\Ecum(\G)\geq-\frac{U}{4}\trace\left[\sqrt{\g(1-\g)}\right],
\end{equation}
where $\g$ denotes the 1RDM corresponding to $\G$.
Again, we emphasize that for the illustration in this section, we restrict ourselves to those 1RDMs that are also pure-state $v$-representable.
A detailed derivation of the above inequalities is presented in App.~\ref{app:bound_hub_dim}. Comparing these bounds with the correlation measures defined in Eqs.~\eqref{eq:zeta-sqrt} and \eqref{eq:zeta_non_idem}, the result can be re-expressed as
\begin{equation}\label{eq:Hubbard-Ecum-estimates}
    -\frac{U}{2}\zeta_\mathrm{ni}(\g)\geq\Ecum(\G)\geq-\frac{U}{2}\zeta_\mathrm{sqrt}(\g)
\end{equation}
by using the spectral theorem to switch between $\g$ and its NONs $\{n_i\}$.
Furthermore, it can be shown that the above bounds are tight, i.e., all admissible values of $\zeta_\mathrm{ni}$ can be approached arbitrarily closely by $\Ecum$ via ground states of suitable Hamiltonians in Eq.~\eqref{eq:hub_dim_scopes_1}. The same holds true for the lower bound involving $\zeta_\mathrm{sqrt}$. 

The above result is related to the RDMFT functional by realizing that all ground states $\G$ of Hamiltonians from the respective scope fulfill
\begin{equation}
    \Fc(\g)=\Ecum(\G),
\end{equation}
with $\Fc$ the RDMFT correlation functional as defined in Eq.~\eqref{eq:Fc} and $\G\mapsto\g$.
The aforementioned tightness of the upper bound enables a trivial application of the coarse-graining of $\Fc$ by identifying the measure of non-idempotency $\zeta_\mathrm{ni}$ as the appropriate correlation measure. It follows via Eq.~\eqref{eq:coarse-func} that
\begin{equation}\label{eq:hub_dim_corr_func_approx}
    \tilde\Fc(\zeta_\mathrm{ni})=\sup_{\mathcal{G}(\mathcal{S}_\mathrm{HD}^{\mathbb{R}})\ni\g\mapsto\zeta_\mathrm{ni}}\Fc(\g)=-\frac{U}{2}\zeta_\mathrm{ni}.
\end{equation}
We can directly read off the correlation temperature $\kc=U/2$ in this case.
An approximation for the ground-state energy as in Eq.~\eqref{eq:E0-variation-approx} is then obtained from this functional as
\begin{equation}\label{eq:hub_func_approx}
    \begin{split}
        \tilde E_0
        &=\inf_\g \left\{ \trace[\hat{h}\g]+\Fmf(\g)-\frac{U}{2}\zeta_\mathrm{ni}(\g) \right\}.
    \end{split}
\end{equation}
As discussed in Sec.~\ref{sec:corr-temp}, in order to ensure that our functional approximation remains variational, we must use the same 1RDM domain for variation of the above functional as was used during coarse-graining in Eq.~\eqref{eq:hub_dim_corr_func_approx}. In the case of the Hubbard dimer, the set of pure-state $v$-representable 1RDMs, $\tilde{\mathcal{G}}(\mathcal{S}_\mathrm{HD}^\mathbb{R})$, is known exactly~\cite{liebert23}. When treating more involved systems, however, we recognize that an exact characterization of the $v$-representable domain is not feasible. In this case, the variation of the approximated functional will generally be carried out over the complete set of $N$-representable 1RDMs, $\RDMset$, that is explicitly known. This introduces the possibility of falling below the ground state energy, as Eq.~\eqref{eq:Fc-inequ} can no longer be ensured for all 1RDMs.

Subsequently, we will examine the individual terms in the functional depicted in Eq.~\eqref{eq:hub_func_approx}, in order to derive properties of its variational minimizer in different regimes of $t$ and $V$. To achieve this, we re-express the functional explicitly in terms of the matrix elements of the 1RDM. This will eventually require the choice of an orbital basis. But first, we identify several basis-set independent properties of the 1RDM.
Recalling that the full Hilbert space has been restricted to its singlet sector, we realize that the up and down spin blocks of the 1RDM coincide, so
\begin{equation}
    \g=\g^\uparrow\oplus\g^\downarrow=\tilde\g\oplus\tilde\g,
\end{equation}
where we set $\tilde\g=\g^\uparrow=\g^\downarrow$.
The normalization of the 1RDM implies that $\tilde\g_{11}=1-\tilde\g_{22}$ in any basis. Also, the formula for the natural occupation numbers can be obtained as
\begin{equation}\label{eq:hub_dim_non}
    n_\pm=\frac{1}{2}\left[1\pm\sqrt{1-4\tilde\g_{11}(1-\tilde\g_{11})+4|\tilde\g_{12}|^2}\right].
\end{equation}
By invoking the Hermitian property of the 1RDM, it follows that the radicand is real-valued and positive. Together with the bounds on fermionic occupation numbers, $0\leq n_i\leq1$, this implies
\begin{equation}\label{eq:hub_dim_bound_g12}
    0\leq|\tilde\g_{12}|\leq\sqrt{\tilde\g_{11}(1-\tilde\g_{11})}.
\end{equation}
For later purpose we calculate the non-idempotency measure from Eq.~\eqref{eq:zeta_non_idem},
\begin{equation}\label{eq:HD-zeta-ni}\begin{aligned}
    \zeta_{\rm ni}(\g) &= 2\left[ n_+(1-n_+) + n_-(1-n_-)\right] \\
    &= 4\left(\tilde\g_{11}(1-\tilde\g_{11})-|\tilde\g_{12}|^2\right).
\end{aligned}\end{equation}

From this point on, for the matrix representation of the 1RDM we fix the localized basis. The matrix elements are, thus, computed as
\begin{align}
        &\tilde\g_{11}=\bra{\Psi}\hat{f}^\dagger_{1\uparrow}\hat{f}_{1\uparrow}\ket{\Psi}
        =\bra{\Psi}\hat{f}^\dagger_{1\downarrow}\hat{f}_{1\downarrow}\ket{\Psi},\\
        &\tilde\g_{12}=\bra{\Psi}\hat{f}^\dagger_{1\uparrow}\hat{f}_{2\uparrow}\ket{\Psi}
        =\bra{\Psi}\hat{f}^\dagger_{1\downarrow}\hat{f}_{2\downarrow}\ket{\Psi}.
\end{align}
Entering $\tilde\g_{11}$ and $\tilde\g_{12}$ as defined above into Eq.~\eqref{eq:hub_func_approx} leads to
\begin{align}
        &\trace[\hat{h}\g]=2V\tilde\g_{11}-4t\Re(\tilde\g_{12})-V, \label{eq:mf_func_approx_loc_bas-1}\\
        &\Fmf(\g) = 2U\tilde\g_{11}(\tilde\g_{11}-1)+U, \label{eq:mf_func_approx_loc_bas-2}\\
        &\begin{aligned}
        \tilde\Fc(\zeta_\mathrm{ni}(\g))&=-\frac U 2 \zeta_\mathrm{ni}(\g)\\
        &=2U\tilde\g_{11}(\tilde\g_{11}-1)+2U|\tilde\g_{12}|^2.
        \end{aligned}
\end{align}
In the Hubbard dimer, where $n_+ =1- n_-$, this exactly reproduces the RDMFT functional of \citet{Csanyi2000} known as ``corrected HF''. This functional was applied to the symmetric Hubbard dimer by \citet{disabatino2015}.
By comparing the behavior of the individual terms under variation of $\tilde\g_{11}$ and $\tilde\g_{12}$, we can conclude certain properties of the minimizer for the approximate functional, $(\tilde\g^*_{11},\tilde\g_{12}^*)$. From the constraints on the matrix elements, it follows for the minimizer that $0\leq\tilde\g_{11}^*\leq1$ and
\begin{equation}\label{eq:HD-g12}
    \tilde\g_{12}^*=\begin{cases}
            \sqrt{\tilde{\g}^*_{11}(1-\tilde\g^*_{11})}, &\text{if}\; t/U>\sqrt{\tilde{\g}^*_{11}(1-\tilde\g^*_{11})},\\
            -\sqrt{\tilde{\g}^*_{11}(1-\tilde\g^*_{11})}, &\text{if}\; t/U<-\sqrt{\tilde{\g}^*_{11}(1-\tilde\g^*_{11})},\\
            t/U, &\text{else.}
        \end{cases}
\end{equation}
Entering the corresponding value of $\tilde\g_{12}$ into Eq.~\eqref{eq:HD-zeta-ni} reveals that the minimizing 1RDM is idempotent in the regime $|t|/U>\sqrt{\tilde{\g}^*_{11}(1-\tilde\g^*_{11})}$.

From Lieb's work~\cite{Lieb1977, LiebHF}, it is known that for systems with positive-semidefinite pair interaction, a variation of the functional $\trace[\hat h\g]+\Fmf(\g)$ over all $\g\in\RDMset$ will recover the same variational minimizer as the standard HF method. This allows us to investigate the behavior of the mean-field functional directly.
For this, we realize that the mean-field functional misses only the correlation functional compared to Eq.~\eqref{eq:hub_func_approx}. According to Eq.~\eqref{eq:mf_func_approx_loc_bas-1}-\eqref{eq:mf_func_approx_loc_bas-2}, this means that only one term in the mean-field functional depends on $\tilde\g_{12}$, specifically on $\Re(\tilde\g_{12})$. From its negative sign, it is understood that the functional value is lowered by maximizing $\Re(\tilde\g_{12})$. For the minimizer of the mean-field functional this implies that it saturates the bound in Eq.~\eqref{eq:hub_dim_bound_g12} as $\tilde\g_{12}^*=\sqrt{\tilde\g_{11}^*(1-\tilde\g_{11}^*)}$. Entering this value of $\tilde\g_{12}$ into Eq.~\eqref{eq:hub_dim_non} reveals that the minimizing 1RDM is idempotent, as required for the HF minimizer.
These insights further lead to the conclusion that whenever the variational minimizer of the \emph{full} functional approximation (see Eq.~\eqref{eq:hub_func_approx}) is idempotent, it coincides exactly with the HF solution.

\begin{figure}[ht!]
    \centering
    \includegraphics[width=\columnwidth]{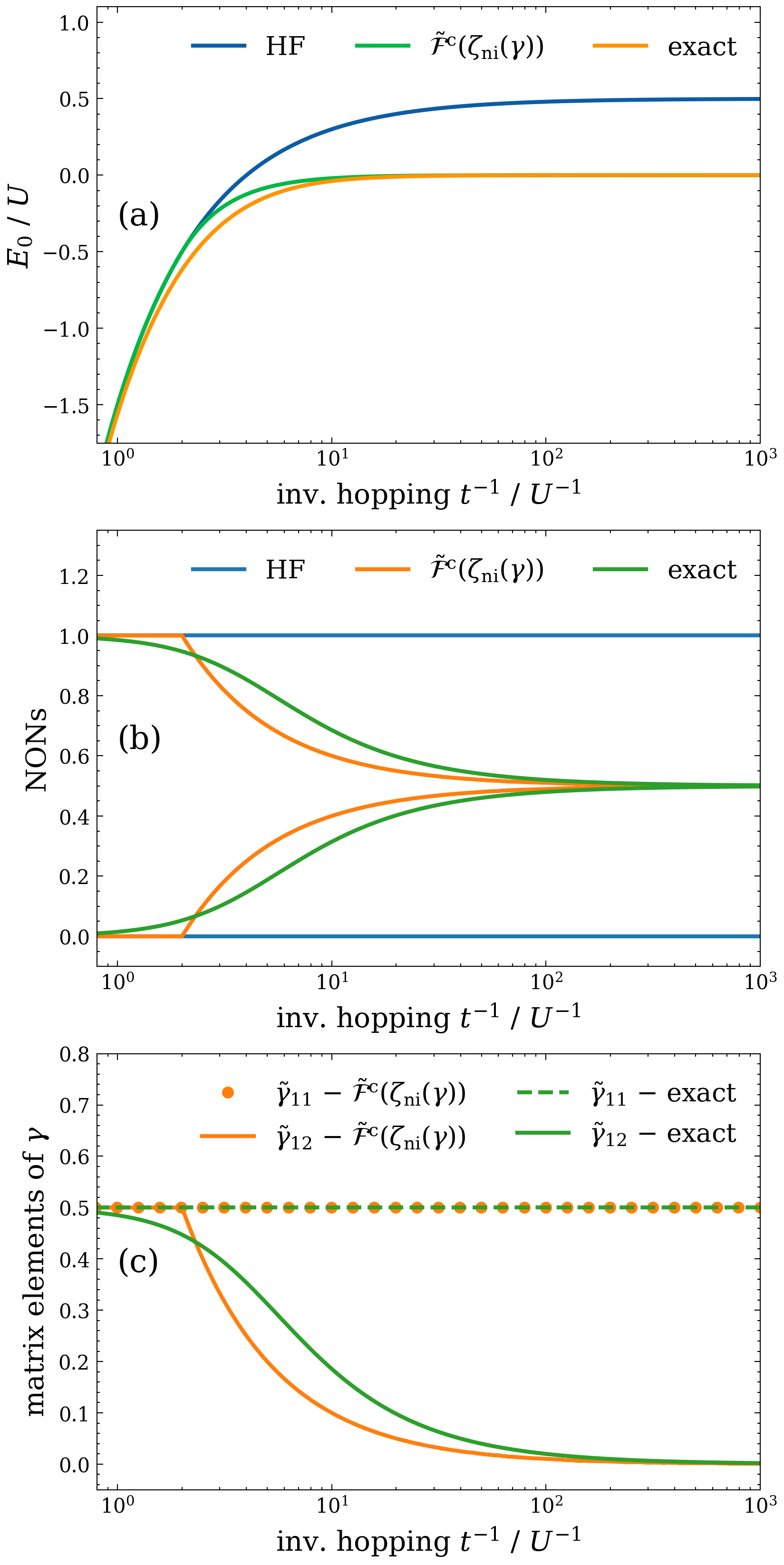}
    \caption{Symmetric Hubbard dimer ($V=0$). Results on (a) ground-state energy, (b) natural occupation numbers, and (c) matrix elements of the 1RDM in the localized basis, over various orders of the inverse hopping strength $t^{-1}$ in the symmetric Hubbard dimer ($V=0$). The results for three different methods are compared: the exact solution to the Hamiltonian (exact), the functional approximation from Eq.~\eqref{eq:hub_dim_corr_func_approx} ($\tilde\Fc(\zeta_\mathrm{ni}(\g))$), and the HF method (HF).
    The interaction strength is fixed to $U=1$.
    }
    \label{fig:hub_dim_V=0}
\end{figure}

Figure~\ref{fig:hub_dim_V=0} presents the results for the RDMFT calculation using the functional approximation $\tilde\Fc(\zeta_\mathrm{ni}(\g))$ in the symmetric Hubbard dimer ($V=0$) over various orders of hopping strength $t$. For this calculation, we restricted the hopping strength to the positive half-space, $t\in\mathbb{R}_0^+$. We established before that the restriction is justified by the particle-hole symmetry of the half-filled system. Recalling that our dimer system can be understood as a minimalistic model of diatomic molecules, the results are plotted against the inverse hopping strength in a semi-logarithmic format. This enables us to relate them to the reduced internuclear distance, scaling as $\log(1/|t|)$ to lowest order. This picture allows for an interpretation of the limit of $1/|t|\to\infty$ as \emph{dissociation} and $1/|t|\to0$ as the \emph{high-density} limit~\cite{LevyPerdew1994}.
The results are compared with the exact solution of the Hubbard dimer and the HF solution. 

The results on the ground-state energy in panel (a) reveal that the fundamental inequality of our approach, Eq.~\eqref{eq:Fc-inequ}, translates, as required, into $E_0\leq \tilde E_0 \leq E_\mathrm{HF}$. It is found that the energy curve from the approximated functional interpolates nicely between the HF energy and the correct ground-state energy. In particular, for $U/t>10$ towards the dissociation limit, the minimization of the approximated functional agrees well with the exact ground-state energy. In the high-density domain, on the other hand, the energy approximation approaches the HF energy curve and even switches to \emph{exactly} the HF result from point of idempotency noted after Eq.~\eqref{eq:HD-g12} onward.

This behavior is mirrored in the evolution of the NONs displayed in panel (b). While the results of the approximated functional agree with the exact solution in both the dissociation and high-density limits, a significant deviation is observed for intermediate values. Again we notice that for $1/t$ below the switching point, the minimizing 1RDM is idempotent, such that the solution of the functional approximation agrees exactly with the HF result.

The same holds true for the evolution of the matrix elements of the 1RDM in the localized basis as of the results in panel (c). The agreement between the matrix elements from our functional approximation and the exact solution of the Hamiltonian, when expressed in the same basis, confirms that a reasonable approximation to the whole 1RDM is also recovered. This is to say that beyond the NONs, the NOs are estimated well too.

\subsection{Symmetric scope \texorpdfstring{$\mathcal{S}_\mathrm{HD}^\mathrm{sym}$}{SDHsym}}\label{sec:hub_dim_symmetric}

Moving on to the symmetric scope $\mathcal{S}_\mathrm{HD}^\mathrm{sym}$ from Eq.~\eqref{eq:hub_dim_scopes_2}, we illustrate how an appropriate treatment of known symmetries within the class of relevant systems can lead to a substantial improvement of the functional approximation. In the present case, it even yields the exact correlation functional in the respective regime.
This is because the correlation functional can be expressed exactly in terms of a correlation measure. In particular, it is found for any Hamiltonian in $\mathcal{S}_\mathrm{HD}^\mathrm{sym}$ that its ground state $\G$ with corresponding 1RDM $\g$ fulfills
\begin{equation}\label{eq:Hubbard-Ecum-sqrt}
    \Ecum(\G)=\Fc(\g)=-\frac{U}{2}\zeta_\mathrm{sqrt}(\g).
\end{equation}
A detailed derivation of this relationship is presented in App.~\ref{app:ana_sol_hub_dim}.
Naturally, the correlation measure with the square-root characteristic is then identified as an appropriate choice for the coarse-graining variable and it follows that
\begin{equation}
    \tilde\Fc(\zeta_\mathrm{sqrt})=\sup_{\mathcal{G}(\mathcal{S}_\mathrm{HD}^\mathrm{sym})\ni\g\mapsto\zeta_\mathrm{sqrt}}\Fc(\g)=-\frac{U}{2}\zeta_\mathrm{sqrt}.
\end{equation}
Hence, a scenario occurs where the functional approximation even coincides with the exact functional,
\begin{equation}
    \F(\g)=\Fmf(\g)+\tilde\Fc(\zeta_\mathrm{sqrt}(\g))\leq\Fmf(\g),
\end{equation}
where again we have correlation temperature $\kc=U/2$.
This means that the results obtained through this functional approximation will agree perfectly with the exact solution of the symmetric Hubbard Hamiltonian.
It is further interesting to note that since in the Hubbard dimer $n_+ =1-n_-$ this exactly reproduces the functionals of \citet{M84} and equally that of \citet{Goedecker1998}.

While in more complex systems, identifying the exact relationship between the true correlation functional and some correlation measure (if there is one) is not feasible, we showed here that the introduction of more information on the characterization of the $v$-representable states for a given scope can result in an improvement of the functional approximation.

\section{Numerical investigations}
\label{sec:num}

The possibility of finding the exact RDMFT correlation functional for a small enough scope, the derivation of the functional's parameters from information about the Hamiltonian, and the estimation of errors of an approximate correlation functional are very attractive for quantum chemistry. They represent a chance to tackle the electronic correlation problem that is often only solved in an imbalanced way: While active-space methods that explicitly diagonalize a many-body Hamiltonian typically describe strong correlation naturally, dynamic correlation requires active spaces beyond the capabilities of current computational hardware. On the other hand, Kohn-Sham DFT approaches use a mean-field approximation within which they augment the one-particle Hamiltonian with a correlation functional that includes the total electron density, but often still fare poorly for strongly correlated multideterminantal systems. A main challenge for RDMFT is the large number of parameters in quantum chemical Hamiltonians and the intricate structure of their minimizers. In this section, we evaluate the connection between correlation energy and entropy for several systems and explore directions to improve the \iDMFT{} method.

\subsection{\texorpdfstring{H$_4^{2+}$}{H4(2+)}: are two-electron systems always simple?}
\label{sec:H4}

In our previous work~\cite{Graf2026}, we studied the dependence of the cumulant energy (Eq.~\eqref{eq:def_e_cum}) on the particle-hole symmetrized entropy for several quantum chemical examples. This was done using high-level methods along molecular deformations, each equivalent to a scope of 1-particle Hamiltonians $\mathcal{S}$, by computing the electronic ground states to form $\mathcal{G}(\mathcal{S})$. We then visualized the distribution of tuples ($\Sph$,$\Ecum$) in these sets. In many cases, we observed that the resulting distributions indeed follow linear $\Sph$-$\Ecum$ curves in spirit of the `correlation conjecture' and such as described by Eq.~\eqref{eq:tildeFc-linear}, just as assumed in \iDMFT{}. This holds for different bond types and molecular deformations, but we identified several requirements for the electronic-structure change during the molecular deformation for this to hold true. Most importantly, the redistribution of electrons may only occur within pairs of orbitals (minimal orbital pairing) that do not change their character. This is typical for many covalent bonds that give rise to one bonding and one antibonding orbital, but not for electron-deficient compounds such as B$_2$ or C$_2$ or heterolytic bond dissociations such as in HeH$^+$. 

In this section, we investigate planar H$_4^{2+}$ in its singlet state, a very simplistic but interesting example for a system with only two electrons where the electronic structure can be described with minimal orbital pairing only along some of the deformation coordinates. The four atoms are weakly bound together by the electron pair, leading to a local energy minimum at square geometry (D$_{4\mathrm{h}}$ symmetry). The lowest-energy A$_{1g}$ orbital is doubly occupied. 

\begin{figure}
    \centering
    \includegraphics[width=\linewidth]{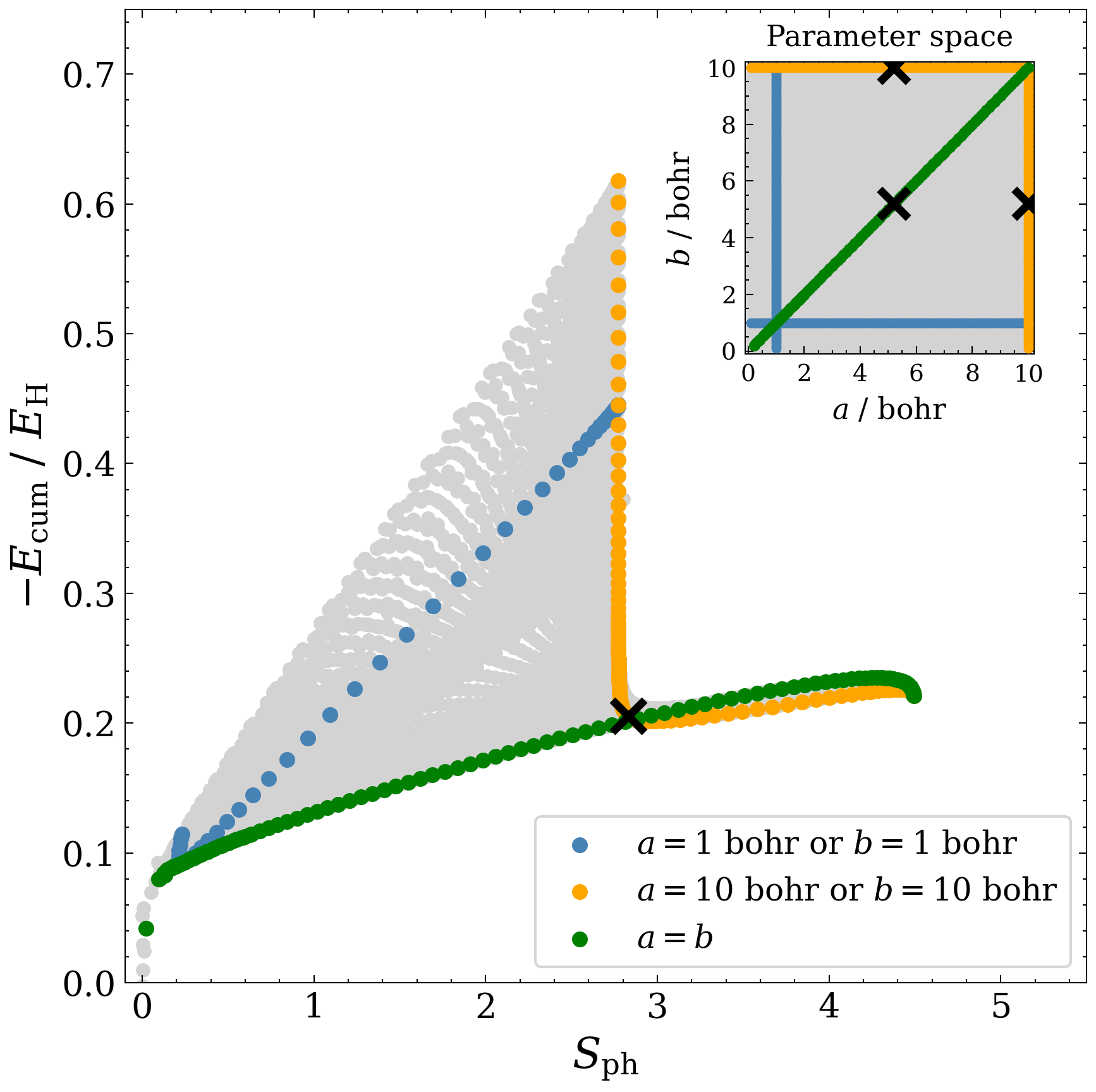}
    \caption{Negative cumulant energy vs. particle-hole symmetrized entropy along different deformations of H$_4^{2+}$, computed with CASSCF(2,20)/cc-pV5Z.}
    \label{fig:h4-diss}
\end{figure}

In our investigations, we carried out CASSCF(2,20)/\allowbreak{}cc-pV5Z calculations using the PySCF software package~\cite{Sun20}. We previously found that this protocol reproduces FCI curves near the basis set limit for the H$_2$ molecule~\cite{Graf2026}. We studied the ground-state wave function for planar-rectangular arrangements of the four atoms in the D$_{2\mathrm{h}}$ point group, such that the geometry of the system is defined by the two side lengths of the rectangle $a$ and $b$ as parameters. Figure~\ref{fig:h4-diss} shows $(\Sph,\Ecum)$ pairs for parameter values $0.1\;\mathrm{bohr}< a,b < 10\;\mathrm{bohr}$ as gray data points. The resulting structure is composed of a triangular shape whose right boundary coincides with $\Sph = 4\log(2) \approx 2.77$, where the upper right corner reaches up to $\Ecum \approx -0.62\;E_\mathrm{H}$. The remaining value pairs with $\Sph > 4\log(2)$ form a thin `tail', which is a continuation of the lower part of the triangle and only reaches up to $\Ecum \approx -0.23\;E_\mathrm{H}$. We highlighted a few parameter paths corresponding to specific one-dimensional ranges of the parameters $a,b$ in color and discuss them separately in the following.

The symmetric dissociation path with $a=b$ maintains the square-planar geometry. In that case, the point group stays D$_{4\mathrm{h}}$, with the lowest four orbitals of A$_{1g}$, degenerate E$_u$, and B$_{1g}$ symmetry. In the limit of dissociation these four become degenerate and correspond to symmetry-adapted representations of electrons localized on each of the nuclei. This is not compatible with the condition of minimal orbital pairing and results in a nonlinear $\Sph$-$\Ecum$-dependence as shown in Fig.~\ref{fig:h4-diss} (green data points). When still fitting Eq.~\eqref{eq:tildeFc-linear} to the resulting values by minimizing the root mean square error of data points weighted by the spacing in $\Sph$, one obtains $\kc=0.0356\;E_\mathrm{H}$ and $b=0.0968\;E_\mathrm{H}$, thus a much lower slope than found for H$_2$ before ($\kc=0.0943\;E_\mathrm{H}$)~\cite{Graf2026}. The data points from the symmetric dissociation also form the lower boundary of the triangle for $\Sph < 4\log(2)$ and the upper boundary of the tail-like structure for $\Sph > 4\log(2)$.

Despite this most straightforward dissociation not following the correlation conjecture (the linear $\Sph$-$\Ecum$ relation), another dissociation mode behaves accordingly because it does obey minimal orbital pairing. When keeping one side length constant at $a=1\;\mathrm{bohr}$, such that there is a pairwise bonding interaction in the molecule, and only varying $b$, two degenerate pairs of orbitals emerge in the limit $b\to\infty$, corresponding to the bonding and antibonding orbitals on the two fragments. The electrons are then only delocalized over the two lowest orbitals, implying that static correlation only happens within this orbital pair. As represented by the blue data points in Fig.~\ref{fig:h4-diss} forming a straight line, the correlation temperature $\kc$ is indeed constant over this dissociation at a value of $0.141\; E_\mathrm{H}$, which is relatively large compared to results for other molecules~\cite{Graf2026}.

In this asymmetric dissociation limit $b\gg a = 1\;\mathrm{bohr}$ 
we find two separated H$_2^+$ systems with one electron each. To connect to the endpoint of the first process ($a=b$ large), one may conceive a third trajectory: changing the internuclear distance $a$ of the two H$_2^+$ fragments pairwise while maintaining them separated (in our calculation, at $b = 10\;\mathrm{bohr}$). As $a$ approaches $b = 10\;\mathrm{bohr}$, the two pairs of degenerate orbitals become almost degenerate as well. The $\Sph$-$\Ecum$ curve of the corresponding ground states exhibits a striking L-shape (orange curve in Fig.~\ref{fig:h4-diss}), which follows the right edge of the triangle and then continues as the lower boundary of the tail structure. This arises from a two-phase dissociation process. First, the single electrons stay localized on the two fragments and occupy the bonding orbital on each site, such that the natural occupation numbers and thus the entropy stays constant. However, the decreasing overlap of these bonding orbitals, and thus the change of the energies of the electrons localized on each fragment H$_2^+$, leads to a change in the value of the cumulant energy. The significant change in their character at constant orbital occupation implies that the correlation conjecture does not hold here, and the correlation effects on the ground-state energy change independently from the entropy. At a distance of around 6 bohr, the entropy starts to increase significantly as the bonding interaction within the pairs becomes so weak that all four orbitals become correlated, which coincides with a slow linear increase.
The whole process resembles a phase transition from a partially frozen state, where the electrons stay localized at their fragments and can thus contain high correlation energies at constant entropy, to a state of electrons delocalized over all four nuclear centers that corresponds more to a metallic phase and has a low and almost constant correlation energy while the entropy increases.

\begin{figure}
    \centering
    \includegraphics[width=\linewidth]{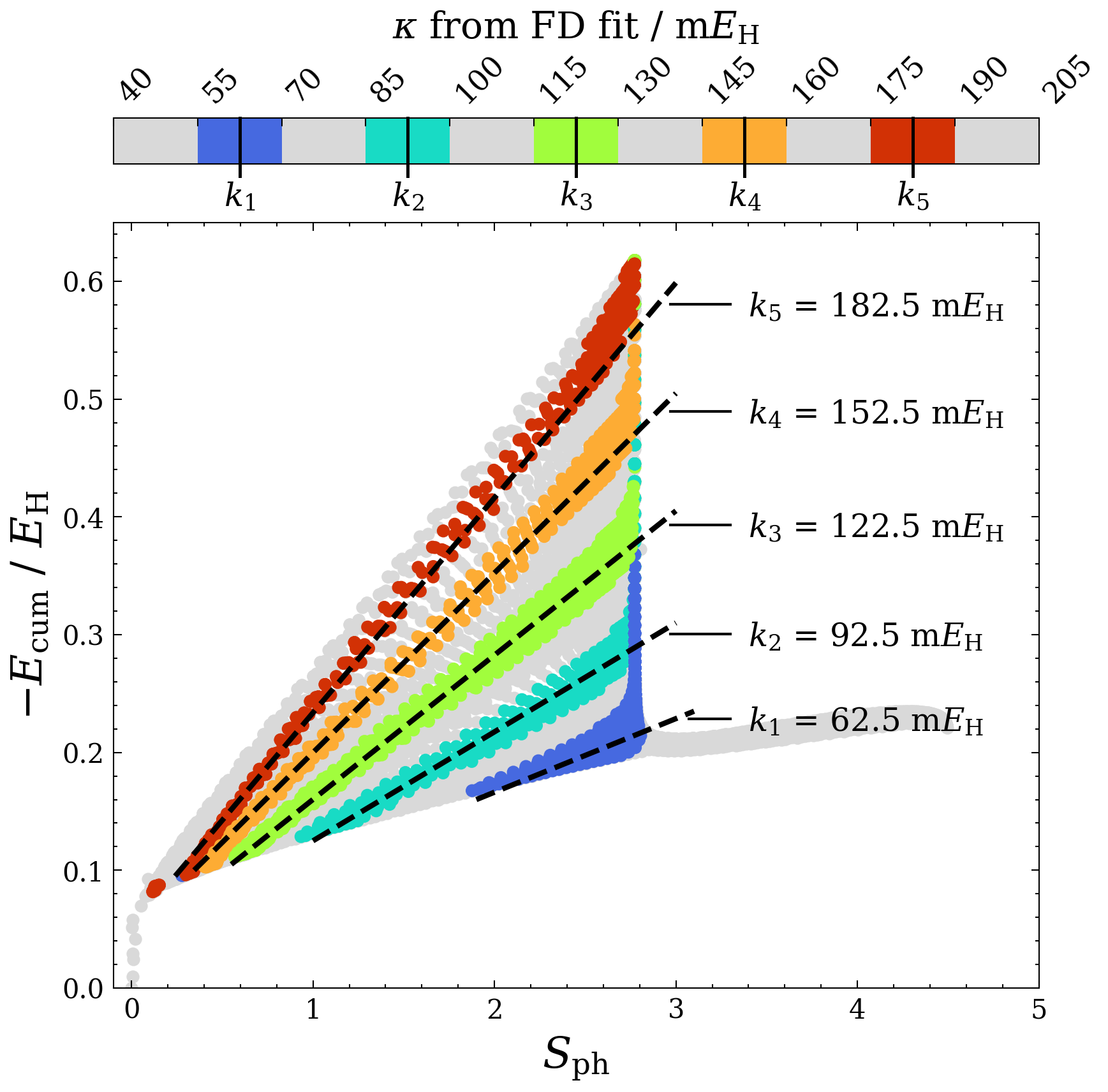}
    \caption{H$_4^{2+}$ computed with CASSCF(2,20)/cc-pV5Z. Negative cumulant energy vs. particle-hole symmetrized entropy from a calculation over a grid for the bond lengths $a$ and $b$, ranging from 0.1 to 10 bohr with a step size of 0.1 bohr. The colors indicate intervals for the correlation temperature $\kc$, obtained by fitting a Fermi-Dirac distribution to the NONs during variation of the bond lengths while maintaining a rectangular configuration. 
    For illustrative purposes, the data points of each colored bin are compared with dashed lines with slopes $k_X$ matching the central value of the respective bin.}
    \label{fig:h4-diss-heat}
\end{figure}

Deducing an effective electronic temperature in a real system is not only possible by looking at the behavior of $\Ecum$ versus $\Sph$ at different parameter values, but can also be carried out on single ground states. Since the thermal HF reference calculation from Sec.~\ref{sec:corr-temp} has a Fermi-Dirac distribution of orbital energies, one can try and determine $\kc$ by just fitting the Fermi-Dirac distribution to the NONs $\{n_i\}$ obtained from the correlated ground state of a high-level method together with some effective orbital energies. Those will be determined as the expectation values of the Fock operator with respect to the NOs $\{\chi_i\}$,
\begin{equation}
\begin{split}
    e_i(\gamma):&=\bra{\chi_i}\hat{F}(\gamma)\ket{\chi_i}\\
    &=\bra{\chi_i}\hat{h}\ket{\chi_i}+\sum_j n_j\bra{\chi_i \chi_j} \hat W \ket{\chi_i \chi_j}.
\end{split}
\end{equation}
Here, the single-particle Hamiltonian is denoted as $\hat h$ and the electron-repulsion integrals are like in Eq.~\eqref{eq:mfFunc}. A similar procedure has previously been suggested in the framework of TAO-DFT~\cite{Chen2022}, where $\kc$ was determined from the occupation number and energy of just the frontier orbital.

We have carried out such a fitting procedure with all ground-state NONs and orbital energies computed at CASSCF level in the H$_4^{2+}$ example. For each geometry, the root-mean-square deviation of the occupation numbers from the Fermi-Dirac distribution was minimized to obtain values for effective temperature $\kc$ and chemical potential $\mu$. One should keep in mind that since the true NONs do not follow an exact Fermi-Dirac distribution, and since there are instances where multiple or all of the significant occupation numbers are almost degenerate, the obtained $\kc$ values come with an uncertainty. 
In Fig.~\ref{fig:h4-diss-heat}, we highlighted data points in the $\Sph$-$\Ecum$ plot where the obtained $\kc$ fall into specific ranges.
The resulting colored regions correspond to linear stripes, similar to what was found before for variation of one side length of the rectangle while keeping the other constant at a small value (the dissociation paths that were shown before to follow the correlation conjecture of  Eq.~\eqref{eq:tildeFc-linear}).
Notably, the approximate inclinations of these stripes read off from the plot correspond to the center of the respective bins in all cases. The concept of correlation temperature thus does not only manifest as a proxy for cumulant energy, but also shows up here in the distribution of the correct NONs---with values matching to a high precision. We stress that this matching effective temperature of Fermi-Dirac-like fits of the NONs is not at all an automatic result, but in turn it justifies the use of a coarse-grained functional that employs correlation temperature and thermal HF as an approximation for the ground-state energy and NONs~\cite{martinez2023}.
Along the other dissociation paths discussed above, i.e., the symmetric dissociation and the variation of one side length while the other is already large, the temperatures from the Fermi-Dirac fit are not constant and do not approximate the slope in the $\Sph$-$\Ecum$ curve. For the symmetric dissociation, $\kc$ determined by both methods decreases slowly, but they do not quantitatively agree. The increase of the second side length when the first side length is already large would first correspond to a divergent correlation temperature when read off from the $\Sph$-$\Ecum$ curve. It returns to a positive value upon the observed turning point, while the temperature from a Fermi-Dirac fit is positive and decreasing throughout. To conclude, one can presume that in molecular systems only some deformations are \emph{distinguished} through an agreement between the electronic temperature in the correlation functional and the effective temperature that reproduces the correct occupation numbers, while for others there is a qualitative disagreement between the two.

\subsection{\texorpdfstring{H$_2$}{H2}: combining correlation measures to describe dynamic correlation}
\label{sec:combining-measures}

In most multireference systems, the biggest contributors to the particle-hole symmetrized entropy as a correlation measure are occupation numbers that significantly differ from zero and full occupation. Such fractional occupations result from static correlation. However, small but nonzero occupation numbers arise in every system with electron-electron interaction from so-called dynamic correlation~\cite{benavidesriveros17}. Even though the error arising from neglecting these small occupation numbers, e.g., when using a single-determinant wave function, is smaller than the error from neglecting static correlation, they still have significant contributions to the total energy. While many of these dynamic correlation contributions do not change much during the breaking of a single bond, their correct estimation is still necessary to compute electronic energy differences with chemical accuracy.

\begin{figure}
    \centering
    \includegraphics[width=0.95\linewidth]{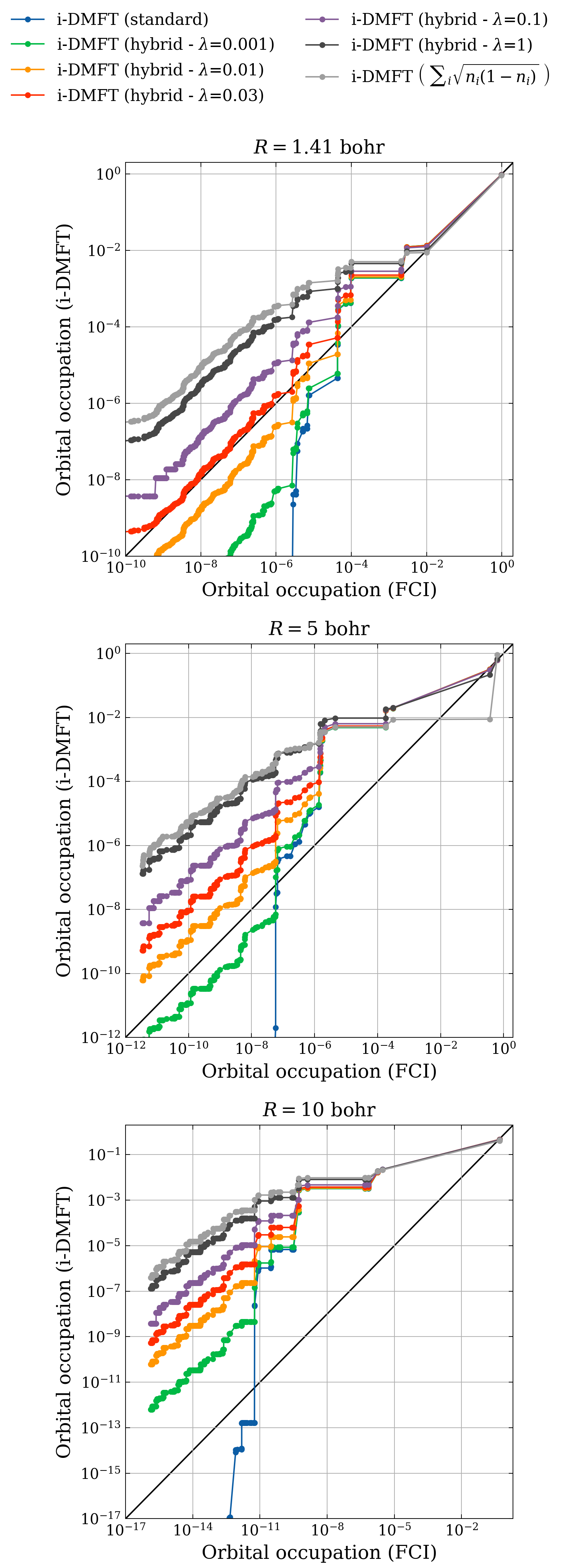}
    \caption{Natural orbital occupations from \iDMFT{} with different entropy-like functionals vs.\ natural orbital occupations from FCI in dihydrogen for three different inter-nuclear distances in cc-pV8Z basis.}\label{fig:NONvsNON}
\end{figure}

Here, we study the set of occupation numbers of a simple and well-understood example: the H$_2$ molecule at three different internuclear distances. It is known that in the basis-set limit, there are infinitely many nonzero occupation numbers, which decay algebraically when ordered by magnitude~\cite{friesecke2003,Giesbertz2013,sobolev2022}.  To approach this limit, we utilize the cc-pV8Z basis set, which has 204 basis functions per H atom, and run Full Configuration Interaction calculations. We also compare these results with \iDMFT{} calculations using different expressions for the correlation measure. Besides the established method that uses the particle-hole symmetrized entropy (Eq.~\eqref{eq:nf}), we also use the measure $\zeta_{\rm sqrt}$ that contains a sum over $\sqrt{n_i(1-n_i)}$, which reproduces the correct boundary behavior at values close to 0 and 1 as mentioned after Eq.~\eqref{eq:zeta-sqrt}. We can also interpolate between the two measures by introducing a weighted sum of the two functions
\begin{equation}\label{eq:hybrid_functional_nf_sqrt}
    \zeta_\mathrm{hybrid}^\lambda(\gamma) = \frac{1}{1+\lambda} (\zeta_\mathrm{nf}(\gamma)+4\log(2)\lambda\zeta_\mathrm{sqrt}(\gamma))
\end{equation}
with $\lambda \geq 0$. The rescaling ensures that both correlation measures have the same maximum value.
The above interpolation is equal to $\zeta_\mathrm{nf}$ when $\lambda=0$ and to $4\log(2)\zeta_\mathrm{sqrt}$ when $\lambda\to\infty$. For different values of $\lambda$ the $\kc$ was chosen to reproduce the reference dissociation energy up to $1\;\upmu E_\mathrm{H}$, equivalent to \iDMFT{} applications in the literature~\cite{WangKnowles21-PRA,WangBaerends22-PRL,Graf2026}.  
The occupation numbers have been ordered by magnitude and are plotted against the reference results in Fig.~\ref{fig:NONvsNON}.
The plots show that occupation numbers computed with the \iDMFT{} functional, i.e., those resulting from a Fermi-Dirac distribution over orbital energies, decrease far too quickly compared to the FCI reference and only very large occupation numbers are captured correctly. Assuming equal spacing in the orbital energies, they decay exponentially, compared to the algebraically decaying FCI results. However, already a small contribution from the square-root functional slows down the decay substantially such that the smallest occupation number is around $10^{-12}$. While the pure square-root functional overestimates the FCI occupation numbers by a constant factor, reducing the $\lambda$ coefficient can align the numbers over a large range. For the equilibrium distance 1.41 bohr, this coefficient is around 0.03; it is between 0.001 and 0.1 for 5 bohr distance and lower than 0.001 at 10 bohr. 

The reason for this difference in $\lambda$ is a general overestimation of many occupation numbers at larger bond lengths. In the exact calculation, these numbers are very small because the system consists of two electrons that are correlated with each other due to entanglement, but are not subject to any further (dynamic) correlation as they are spatially separated. However, \iDMFT{} calculation cannot capture this as the same distribution of occupations over energy levels applies independent of the extent of the electrons' local interaction with each other. Both the particle-hole symmetrized entropy and the $\zeta_\mathrm{sqrt}$ correlation measure thus suffer from self-interaction, which is why occupation numbers are overestimated. While the decay behavior of small occupation numbers can be corrected in the limit, the larger ones do not depend much on the functional choice. The range in which the modified functional can correct the occupation numbers shrinks towards the dissociation limit.

\section{Summary and outlook}
\label{sec:summary}

This perspective started from the common wisdom that a better understanding of how to quantify electron correlation should lead to more efficient treatments of the correlation problem. We have substantiated this view and turned it into a concrete framework for the ground-state problem. Its key implication is a change in the role of correlation measures: so far used mainly as diagnostic tools, they become constructive ingredients of computational methods. Since prominent measures such as the nonfreeness depend only on the 1RDM, the framework naturally resides within RDMFT: for a chosen scope of systems, the correlation functional is coarse-grained into a functional of a single measure, and the resulting self-consistent equations retain the cost of HF. The variable conjugate to the measure then sets the scale on which the occupation numbers are distributed over the orbital energies, much like a temperature. 
For the nonfreeness, which for pure states equals the entropy of the closest mean-field state, the occupation numbers follow a Fermi--Dirac distribution and, for a fixed value of this variable, the equations become identical to those of thermal HF. The correlation temperature thus remains an auxiliary quantity, but one that follows from quantifying electron correlation rather than being introduced ad hoc.

Correlation measures, RDMFT, and effective temperatures thereby emerge as facets of a single picture with a unifying character. Approaches developed largely independently turn out to be rungs of one ladder of approximations: constant correlation corrections on top of HF~\cite{Melius2000,Witkowski2025} and the linear dependence on the nonfreeness employed in \iDMFT{}~\cite{WangBaerends22-PRL} are examples of its lowest rungs, and related approaches such as TAO-DFT~\cite{Chai2012,Chai2014,Yeh2022} and entropy-corrected pair-density functional theory~\cite{Feng2025} can be viewed in the same light. Identifying this ladder sets a clear agenda, which we propose here: to climb it systematically, all at mean-field cost. Moreover, the scope becomes a new design parameter: giving up universality yields approximations that are bounded by the exact and the HF functional within the scope and improve as the scope narrows.

Our examples illustrate these concepts rather than benchmark a method. The analytically solvable Hubbard dimer shows how the scope determines a suitable measure, which for the symmetric dimer even yields the exact functional. For H$_4^{2+}$, a single correlation temperature describes deformations in which the electrons redistribute within fixed orbital pairs, consistent with Fermi--Dirac fits to the exact occupation numbers, but not deformations that change the character of the orbitals. For H$_2$, the behavior of the measure close to full and zero occupation controls the decay of the natural occupation numbers and hence the description of dynamic correlation: adding a square-root contribution with the correct boundary behavior~\cite{schilling2019b,Maciazek2021,Wang2026} substantially improves on the exponential decay produced by the nonfreeness alone. The choice of the measure thus directly shapes the quality of the description, just as the initial wisdom suggests.

Three directions for future work follow. First, the constant correlation temperature of \iDMFT{} corresponds to the simplest, linear approximation; higher rungs require temperatures that vary with the measure~\cite{Hu2024-dispersion} or draw on further information from the 1RDM, for which self-consistent schemes from TAO-DFT~\cite{Lin2017,Chen2022} and estimates from electron repulsion integrals~\cite{Gibney2023} offer starting points. Second, the freedom in choosing the measure invites a systematic search for measures, or combinations thereof, that respect exact conditions and capture static and dynamic correlation alike. Third, since a single scalar cannot follow every change in the entanglement pattern of a molecule, principled choices of scopes or the use of several measures at once will be needed.

Beyond these steps, we advocate a broader shift in how the correlation problem is approached. Rather than asking only how much accuracy can be gained by investing more computational resources, we ask how much of the correlation problem, in particular for strongly correlated systems, can be solved at mean-field cost.
Our framework goes well beyond posing this question: it provides the concepts, the hierarchy of approximations, and first illustrations needed to answer it systematically, with correlation measures as its central design principle. We hope that this perspective inspires further work at the interface of quantum information theory and quantum chemistry.

\section*{Statement on the use of generative AI}
The authors used Claude Opus 5.5 (Anthropic) to help polish the language of the manuscript, in particular the abstract, introduction, and summary and outlook, through iterative revisions based on the authors' content and feedback. All AI-assisted text was reviewed and verified by the authors, who take full responsibility for the content.

\begin{acknowledgments}
We thank D.~Aliverti-Piuri, L.~Ding, and J.~Liebert for helpful discussions. We acknowledge financial support from the German Research Foundation (Grant SCHI 1476/1-1) (P.G.G., F.M., M.P., C.S.), the ERC-2021-STG under grant agreement No.~101041487 REGAL (M.P.), and the ERC-2025-COG under grant agreement No.~101231616 beyondDFT (C.S.). 
\end{acknowledgments}

\appendix

\section{Bounds for the cumulant energy in the Hubbard dimer}\label{app:bound_hub_dim}

This section derives the bounds of the ground-state cumulant energy of the Hubbard dimer as presented in Eq.~\eqref{eq:hub_e_cum_bound}. Their validity will be shown for ground-states of Hamiltonians from the full real-valued scope $\mathcal{S}_\mathrm{HD}^\mathbb{R}$.

Since we consider only pure states in the singlet sector of the Hilbert space for the ground-state search of the Hubbard dimer, all ground states assume the following form:
\begin{equation}\label{eq:app_gs}
    \ket{\Psi}=a_1\ket{\Phi_1}+a_2\ket{\Phi_2}+a_3\ket{\Phi_3},
\end{equation}
with coefficients $a_1,a_2,a_3\in\mathbb{C}$.
The matrix elements of the spin-independent 1RDM $\tilde\g$ ($=\g^\uparrow=\g^\downarrow$) in the localized basis follows as
\begin{equation}
    \begin{split}
        &\tilde\g_{11}=\omega_1+\frac{1}{2}\omega_3,\\
        &\tilde\g_{12}=\frac{1}{\sqrt{2}}(a^*_1a_3+a_2a^*_3),
    \end{split}
\end{equation}
where $\omega_i:=|a_i|^2$ and $^*$ denotes the complex conjugate. From here on, we fix the global phase such that $a_3\in\mathbb{R}^+_0$. The NONs are computed as
\begin{equation}
    n_\pm=\frac{1}{2}\left[1\pm\sqrt{1-\big|\omega_3-2a_1a_2\big|^2}\right].
\end{equation}

Critically, it is observed that the Hubbard interaction is diagonal in the localized basis, which implies that the cumulant energy is fully determined by the diagonal elements of the 2RDM cumulant,
\begin{equation}
    \lambda_2=\G^{(2)}-\frac{1}{2}(\hat{I}-\hat{X})\g\otimes\g,
\end{equation}
with $\G^{(2)}$ denoting the 2RDM of our $N$-particle state.
This enables the derivation of properties of the cumulant energy without explicitly diagonalizing the Hamiltonian. Concretely, the diagonal elements of $\lambda_2$ are obtained as \cite{kong2011}
\begin{equation}
    (\lambda_2)^{a\sigma,b\tilde\sigma}_{a\sigma,b\tilde\sigma}=\mathrm{cov}(a\sigma,b\tilde\sigma)+\g^{a\sigma}_{b\tilde\sigma}\g_{a\sigma}^{b\tilde\sigma}
\end{equation}
with $a,b$ indexing spatial orbitals and $\sigma,\tilde\sigma$ indexing spin states. As of Ref.~\cite{kong2011}, the function $\mathrm{cov}$ is defined as
\begin{equation}
    \begin{split}
        \mathrm{cov}(a\sigma,b\tilde\sigma)&=\sum_\mu \omega_\mu n_{a\sigma}(\mu)n_{b\tilde\sigma}(\mu)\\
        &-\Big(\sum_\mu\omega_\mu n_{a\sigma}(\mu)\Big)
        \Big(\sum_\mu\omega_\mu n_{b\tilde\sigma}(\mu)\Big),
    \end{split}
\end{equation}
where $\mu$ runs over Slater determinants in the CI expansion of the underlying state, expressed in the localized basis. The modulus squared of the corresponding CI coefficients is denoted as $\omega_\mu$. The parameters $n_{a\sigma}(\mu)$ and $n_{b\tilde\sigma}(\mu)$ assume the value $1$ or $0$, depending on whether the respective spin orbital is occupied in the Slater determinant indexed by $\mu$.
For the Hubbard dimer, this leads to
\begin{equation}\label{eq:app_e_cum_gs}
    \Ecum(\Psi)=-\frac{U}{2}(\omega_3^2-4\omega_1\omega_2).
\end{equation}
From the definition of the cumulant energy in Eq.~\eqref{eq:def_e_cum}, it follows that the cumulant energy for ground states is bounded from above by $0$. This means it must hold $\omega_3^2\geq 4\omega_1\omega_2$ and thus $\omega_3 \geq 2|a_1||a_2|$.

On the other hand, from the formula for the NONs, the correlation measures $\zeta_\mathrm{ni}$ and $\zeta_\mathrm{sqrt}$ can be re-expressed as
\begin{equation}\label{eq:app_corr_meas}
    \begin{split}
        &\zeta_\mathrm{sqrt}(\g)=\frac{1}{2}\trace\left[\sqrt{\g(1-\g)}\right]=\big|\omega_3-2a_1a_2\big|,\\
        &\zeta_\mathrm{ni}(\g)=\trace\left[\g(1-\g)\right]=\big|\omega_3-2a_1a_2\big|^2.\\
    \end{split}
\end{equation}

\subsection{Lower bound of \texorpdfstring{$\Ecum$}{Ecum}}

The lower bound for $\Ecum$ is obtained by realizing that $\omega_1+\omega_2\geq2|a_1||a_2|$, which follows from $(|a_1|-|a_2|)^2\geq0$,
\begin{equation}\label{eq:app_lower_bound}
    \begin{split}
        \Ecum(\Psi)&=
        -\frac{U}{2}(\omega_3-2|a_1||a_2|)\cdot(\omega_3+2|a_1||a_2|)\\
        &\geq-\frac{U}{2}(\omega_3-2|a_1||a_2|)\cdot\underbrace{(\omega_1+\omega_2+\omega_3)}_{=1}\\[-0.9em]
        &\geq-\frac{U}{2}|\omega_3-2a_1a_2|.
    \end{split}
\end{equation}
In the last step the reverse triangle inequality was employed.
Comparing this result to Eq.~\eqref{eq:app_corr_meas}, reveals that
\begin{equation}
    \Ecum(\Psi)\geq-\frac{U}{2}\zeta_\mathrm{sqrt}(\g).
\end{equation}
In particular, from Eq.~\eqref{eq:app_lower_bound} it follows that the bound is saturated exactly when $a_1=a_2\in\mathbb{R}$. This is the case for the symmetric dimer, where no site is energetically favorable.

\subsection{Upper bound of \texorpdfstring{$\Ecum$}{Ecum}}

The upper bound for $\Ecum$ is derived from the following inequality involving the non-idempotency correlation measure $\zeta_\mathrm{ni}$: by fixing the phase of $a_3\in\mathbb{R}_0^+$, and with $\varphi$ denoting the cumulated phase of $a_1$ and $a_2$ (i.e., $a_1a_2=e^{i\varphi}|a_1||a_2|$), we obtain for $\varphi\in[-\frac{\pi}{2},\frac{\pi}{2}]$ that
\begin{equation}
    \begin{split}
        \zeta_\mathrm{ni}(\g)&=\left|\omega_3-2a_1 a_2\right|^2\\
        &=\omega_3^2-4|a_1||a_2|\big(\omega_3\cos(\varphi)-|a_1||a_2|\big)\\
        &\leq            \omega_3^2-4\omega_1\omega_2(2\cos(\varphi)-1).
    \end{split}
\end{equation}
Here, we used again that $\omega_3\geq2|a_1||a_2|$ to replace the $\omega_3$ in the term with $\cos(\varphi)$.
Comparing this to the formula of the cumulant energy, it is found that 
\begin{equation}\label{eq:app_upper_bound_zeta}
    \Ecum(\Psi)\leq-\frac{U}{2}\zeta_\mathrm{ni}(\g),
\end{equation}
which holds for all ground states exactly when $\varphi=0$.

Finally, it remains to be shown that the cumulated phase of $a_1$ and $a_2$ vanishes for ground states in the scope $\mathcal{S}_\mathrm{HD}^\mathbb{R}$. Remember that we fix $a_3\in\mathbb{R}_0^+$ and let us denote the phases of $a_1$ and $a_2$ as $\alpha$ and $\beta$, respectively. The expectation value of the Hamiltonian for a trial singlet state (complying with the gauge $a_3\in\mathbb{R}^+_0$) is computed as
\begin{equation}
    \begin{split}
        &\bra{\Psi}\hat{H}_\mathrm{HD}\ket{\Psi}
        =(U+V)\omega_1+(U-V)\omega_2\\
        &\qquad\qquad\qquad-t2\sqrt{2}|a_3|\Big[|a_1|\cos(\alpha)+|a_2|\cos(\beta)\Big]\\
        &\geq (U+V)\omega_1+(U-V)\omega_2-|t|2\sqrt{2}|a_3|\big[|a_1|+|a_2|\big].
    \end{split}
\end{equation}
Observing that the lower bound is saturated when $\alpha$ and $\beta$ are both equal to either 0 ($t>0$) or $\pi$ ($t<0$) entails that the cumulated phase of $a_1$ and $a_2$ vanishes for ground states of the scope $\mathcal{S}_\mathrm{HD}^\mathbb{R}$. This confirms the validity of the bound derived in Eq.~\eqref{eq:app_upper_bound_zeta}.

\section{Solution of the symmetric Hubbard dimer (\texorpdfstring{$V=0$}{V=0})}
\label{app:ana_sol_hub_dim}

This section presents the explicit analytical solution of the symmetric Hubbard dimer. As stated in Sec.~\ref{sec:Hubbard}, the ground-state of the Hubbard dimer, governed by Eq.~\eqref{eq:hub_ham}, has support on the singlet sector of the two-particle Hilbert space. Furthermore, the symmetric property of the system discussed here allows for a further restriction of the Hilbert space to the \emph{gerade} subspace. The elements of this subspace remain fully invariant under exchange of the site labels $1\leftrightarrow2$. To exploit this, first, we introduce the symmetry-adapted single-particle states:
\begin{equation}
    \ket{g}=\frac{1}{\sqrt{2}}\big(\ket{1}+\ket{2}\big),\qquad\ket{u}=\frac{1}{\sqrt{2}}\big(\ket{1}-\ket{2}\big),
\end{equation}
where $\ket1$ and $\ket2$ denote the elements of the localized site basis.
The newly found $\ket g$ and $\ket u$ are eigenstates of the $C_2$ symmetry operation. The labels for the states $g$ and $u$ are inspired by the symbols conventionally used to denote \emph{gerade} and \emph{ungerade} molecular orbitals in the study of electronic structure.

An orthonormal basis for the two-particle singlet Hilbert space in terms of the symmetry-adapted states is obtained as
\begin{equation}
    \begin{split}
        &\ket{\Sigma_1}=\ket{\Phi_1}/2+\ket{\Phi_2}/2+\ket{\Phi_3}/\sqrt{2},\\
        &\ket{\Sigma_2}=\ket{\Phi_1}/2+\ket{\Phi_2}/2-\ket{\Phi_3}/\sqrt{2},\\
        &\ket{\Sigma_3}=\ket{\Phi_1}/\sqrt2-\ket{\Phi_2}/\sqrt2,
    \end{split}
\end{equation}
where the states $\ket{\Phi_i}$ are defined in Eq.~\eqref{eq:hub_dim_singlet_onb}. The first two states in the above span the gerade subspace of the two-particle Hilbert space, since they remain fully invariant under exchange of the labels $1\leftrightarrow2$. In contrast, $\ket{\Sigma_3}$ changes sign under the exchange and therefore spans the ungerade subspace.

As mentioned above, for the ground-state search at $V=0$, we can restrict the set of states to the singlet, gerade subspace.
Projecting the Hamiltonian into this space, which is spanned by $\ket{\Sigma_1}$ and $\ket{\Sigma_2}$, yields
\begin{equation}
    H^{\mathrm{g},S=0}_\mathrm{HD}=
    \begin{bmatrix}
        -2t+U/2 & U/2\\
        U/2 & 2t+U/2
    \end{bmatrix}.
\end{equation}
Here, the superscript indicates the restriction to the gerade ($\mathrm{g}$) and singlet ($S=0$) sector.

The eigenstates of this Hamiltonian are obtained as 
\begin{equation}
    \begin{split}
        \ket\pm=\,&\bigg[2\left(1+\tau^2\right)\pm2\tau\sqrt{1+\tau^2}\bigg]^{-1/2}\\
            &\times\left[\ket{\Sigma_1}+\big(\tau\pm\sqrt{1+\tau^2}\big)\ket{\Sigma_2}\right],
    \end{split}
\end{equation}
with the corresponding eigenvalues
\begin{equation}
    \varepsilon_\pm=\frac{U}{2}\pm\frac{U}{2}\sqrt{1+\tau^2}.
\end{equation}
The parameter $\tau$ introduced in the above is defined as $\tau:=4t/U$. Hence, the ground state is $\ket{-}$. From the singlet property it is known that the projections of the 1RDM into the spin sectors coincide. The NONs are obtained as
\begin{equation}
    \begin{split}
        &n_+=n_+^\uparrow=n_+^\downarrow=\frac{1}{2}\left(1+\frac{|\tau|}{\sqrt{1+\tau^2}}\right),\\
        &n_-=n_-^\uparrow=n_-^\downarrow=\frac{1}{2}\left(1-\frac{|\tau|}{\sqrt{1+\tau^2}}\right).
    \end{split}
\end{equation}
Comparing this with the explicit form of the ground state, it is revealed that $\ket{-}=\sqrt{n_+}\ket{\Sigma_1}-\sqrt{n_-}\ket{\Sigma_2}$ for $t>0$ and $\ket-=\sqrt{n_-}\ket{\Sigma_1}-\sqrt{n_+}\ket{\Sigma_2}$ for $t<0$. Re-expressing the ground state in terms of the localized basis $\{\ket{\Phi_i}\}$ leads to
\begin{equation}
    \ket-=\pm\frac{\sqrt{n_+}-\sqrt{n_-}}{2}\big(\ket{\Phi_1}+\ket{\Phi_2}\big)+\frac{\sqrt{n_+}+\sqrt{n_-}}{\sqrt{2}}\ket{\Phi_3},
\end{equation}
where the sign in front of the first term matches the sign of the hopping strength $t$. This allows us to compute the cumulant energy via Eq.~\eqref{eq:app_e_cum_gs}:
\begin{equation}
    \Ecum=-U\sqrt{n_+n_-}=-\frac{U}{4}\trace\left[\sqrt{\g(1-\g)}\right],
\end{equation}
where the last expression is obtained by realizing that $n_++n_-=1$.

\bibliography{refs}

\end{document}